\documentclass[journal]{IEEEtran}

\usepackage[T1]{fontenc}
\usepackage[utf8]{inputenc}

\usepackage{mathdots}
\usepackage{mathtools}
\usepackage{amsmath}
\usepackage{dsfont}

\usepackage{cite}

\ifCLASSINFOpdf
  \usepackage[pdftex]{graphicx}
  \graphicspath{{./pdf/}{./eps/}{./Figures/}}

  \DeclareGraphicsExtensions{.pdf,.tiff,.eps}
\else
  \usepackage[dvips,ps2pdf]{graphicx}
  \graphicspath{{./pdf/}{./eps/}{./Figures/}}
  \DeclareGraphicsExtensions{.pdf,.tiff,.eps}
\fi
\usepackage[outdir=./pdf/]{epstopdf}

\ifCLASSOPTIONcompsoc
    \usepackage[caption=false, font=normalsize, labelfont=sf, textfont=sf]{subfig}
\else
\usepackage[caption=false, font=footnotesize]{subfig}
\fi

\usepackage{amsmath}
\usepackage[cmintegrals]{newtxmath}

\usepackage{algorithmic}
\usepackage{algorithm}

\usepackage{mathrsfs}

\newcommand{\mypar}[1]{\vspace{0.1in}{\bf #1}}
\newcommand{\myparni}[1]{\vspace{0.1in}\noindent{\bf #1}}

\usepackage{array}

\usepackage{url}
\usepackage[numbers,square]{natbib}

\usepackage{color}
\usepackage[usenames,dvipsnames,svgnames,table]{xcolor}

\usepackage{tikz}
\usetikzlibrary{arrows}
\usetikzlibrary{fit}
\usetikzlibrary{calc}
\usetikzlibrary{circuits.ee.IEC}

\usetikzlibrary{positioning}
\usetikzlibrary{automata,positioning}
\usepackage{pgfplots}

\begin{document}

\IEEEoverridecommandlockouts
\IEEEpubid{\begin{minipage}{\textwidth}\
\\[20pt] 
\centering\normalsize{\copyright 2022 IEEE. DOI: 10.1109/TCOMM.2022.3141741}
\end{minipage}}

\title{SISO Decoding of $\mathbb{Z}_4$ Linear Kerdock and \\ Preparata Codes}

\author{Aleksandar~Minja,~\IEEEmembership{Member,~IEEE,}
        and~Vojin~Šenk,~\IEEEmembership{Member,~IEEE}
\thanks{This work was in part supported by the European Union Horizon 2020 research and innovation program under the WIDESPREAD grant agreement No 856967 and the Serbian Ministry of Education, Science and Technological Development through the project no. 451‑03-68/2020-14/200156: "Innovative scientific and artistic research from the Faculty of Technical Sciences activity domain".}
\thanks{A. Minja is with the University of Novi Sad, Faculty of Engineering (Technical Sciences), Department of Power, Electronic and Telecommunication Engineering, Novi Sad 21000 Serbia e-mail: aminja@uns.ac.rs.}
\thanks{V. Šenk is with the University of Novi Sad, Faculty of Engineering (Technical Sciences), Department of Power, Electronic and Telecommunication Engineering, Novi Sad 21000 Serbia e-mail: vojin\_senk@uns.ac.rs.}
\thanks{Manuscript received March 20, 2021.}}

\markboth{Journal of \LaTeX\ Class Files,~Vol.~14, No.~8, August~2015}%
{Shell \MakeLowercase{\textit{et al.}}: Bare Demo of IEEEtran.cls for IEEE Journals}

\maketitle

\begin{abstract}
Some nonlinear codes, such as Kerdock and Preparata codes, can be represented as binary images under the Gray map of linear codes over rings. 
This paper introduces MAP decoding of Kerdock and Preparata codes by working with their quaternary representation (linear codes over $\mathbb{Z}_4$) with the complexity of $\mathcal{O}(N^2\log_2 N)$, where N is the code length in $\mathbb{Z}_4$. A sub-optimal bitwise APP decoder with good error-correcting performance and complexity of $\mathcal{O}(N\log_2 N)$ that is constructed using the decoder lifting technique is also introduced. This APP decoder extends upon the original lifting decoder by working with likelihoods instead of hard decisions and is not limited to Kerdock and Preparata code families. Simulations show that our novel decoders significantly outperform several popular decoders in terms of error rate.
\end{abstract}


\begin{IEEEkeywords}
Codes over rings, Kerdock codes, MAP decoding, Preparata codes, Quaternary cyclic codes.
\end{IEEEkeywords}

\IEEEpeerreviewmaketitle

\section{Introduction}

\mypar{Context and motivation.} It was shown in \cite{ahammons1} that some families of nonlinear binary codes with good properties can be represented as the binary image under the Gray map of linear codes over $\mathbb{Z}_4$ (the ring of integers modulo 4). These families include the Kerdock \cite{akerdock1}, Preparata \cite{apreparata1}, Goethals \cite{agoethals1,agoethals2} and Delsarte-Goethals \cite{adg1} codes. It is well known that the quaternary Nordstrom-Robinson code \cite{anordstrom1} (also known as the “octacode” \cite{bconway1}, which is used when the Leech lattice is constructed from eight copies of the face-centered cubic lattice \cite{ahammons1}) is a self-dual code of length 8 and represents the initial member of both Kerdock and Preparata families \cite{ahammons1}. 

The decoding of the nonlinear binary codes (and their $\mathbb{Z}_4$ linear counterparts) was an important area of research for years, and many hard and soft decision decoding algorithms were proposed. Almost all soft-decision decoding algorithms are based on an exhaustive search approach with respect to the Euclidean distance or the correlation between the received channel sequence and the possible transmitted codeword. With the exception of the Nordstrom-Robinson, there is no low-complexity soft-input soft-output (SISO) decoder for the mentioned families of nonlinear binary codes. We believe this to be an important obstacle for their adoption in modern coding systems.

The primary purpose of this paper is to fill the void and provide the necessary tools for using these codes in modern coding systems. In this paper, we focus only on Kerdock and Preparata families, but similar techniques as the ones considered here can be used to design decoding algorithms for other related families.

\mypar{Use of nonlinear codes and codes over rings in communication systems.} There is widespread use of the Nordstrom-Robinson code in digital systems \cite{ali1,axie1}. A turbo product coding scheme with Nordstrom-Robinson as a component code was presented in \cite{arocha2,arocha1}. Kerdock codes for limited feedback MIMO systems were presented in \cite{atakao1,atakao2}. Space-time codes for standard phase shift keying (PSK) and quadrature amplitude modulation (QAM) signal constellations based on Kerdock and Delsarte-Goethals codes were presented in \cite{acalderbank1}. Quaternary constant-amplitude codes (codes that reduce the peak-to-average power ratio) for OFDM and MC-CDMA were presented in \cite{adavis1,aschmidt1}. Kerdock coded MC-CDMA system with non-linear amplifiers was presented in \cite{ashakeel1}. More recently, nonlinear codes are often used as components of product codes in order to achieve better reliability and security \cite{aamrani1,akarp1}. Codes over rings have also found application in coding for the phase noise channel \cite{aliva1} and network-coded bidirectional relaying systems \cite{7108018}. 

\mypar{Brief overview of existing decoding algorithms.} Common approaches for designing decoders for codes over rings include \cite{abarrolleta1} the algebraic (syndrome) decoding of linear codes \cite{macwilliams1977theory}, the lifting decoder technique \cite{agreferath1,ababu1}, the partitioning of a code into the subcode and its cosets (a.k.a. the coset decomposition approach), introduced by Conway and Sloane in \cite{aconway1} (which works for both linear and nonlinear codes), and the permutation decoding \cite{abarrolleta1,Barrolleta2018,abarrolleta2}. Most of the soft decision decoders employ some form of the coset decomposition approach. E.g., for the Reed-Muller (RM) \cite{akerdock1} subcode, we can use the fast Hadamard transform (FHT), which can significantly reduce the complexity of the decoder. A low complexity maximum likelihood (ML) decoding of Nordstrom-Robinson code was presented in \cite{aadoul1}. An efficient implementation of this algorithm was presented in \cite{aabbaszadeh1} and an improvement was proposed in \cite{avardy1}. A similar decoding algorithm of the same order of complexity was presented for Kerdock codes in \cite{ahammons1} and \cite{aelia1}. A product code representation of nonlinear codes, using a high-complexity list-based ML decoding, along with a low-complexity bounded distance decoding algorithm based on coset decomposition, was presented in \cite{aamrani1}. It was noted in \cite{aamrani1} that effective decoding of Kerdock codes based on the proposed representation is left as an open problem. The standard way of implementing the maximum a posteriori (MAP) decoding of a linear code is via the Bahl–Cocke–Jelinek–Raviv (BCJR) algorithm \cite{bcjr}, which uses a trellis representation of the code. The trellis complexity of Kerdock and Delsarte-Goethals codes was presented in \cite{ashany1}, where a trellis representation consisting of parallel sections, each corresponding to a different coset of the RM$(1,m)$\footnote{Where RM$(r,m)$ represents the $r$-order RM code of length $2^m$ \cite{macwilliams1977theory}.} subcode was considered. As the complexity of the BCJR algorithm for the RM$(1,m)$ code is $\mathcal{O}(N^2)$ (where $N$ is the length of the code) \cite{aashikhmin1}, the complexity of the BCJR decoding of the corresponding Kerdock code operating on this trellis would be $\mathcal{O}(N^3)$\footnote{It was noted in \cite{ashany1} that the Viterbi decoding on the biproper trellis would improve run-time performance by up to $10\%$, compared with the decoding on the trellis consisting of parallel cosets of the RM$(1,m)$ code.}. The trellis complexity of the Preparata and Goethals codes was investigated in \cite{ashany2}. A twisted squaring trellis construction based on the extended primitive double error correcting BCH code and its cosets was proposed in \cite{ashany2}. This construction is an extension of the Nordstrom-Robinson representation presented in \cite{aforney1}. A trellis representation of the Nordstrom-Robinson code based on the generalized array code was presented in \cite{bhonary1}. A decoding algorithm based on this representation was investigated in \cite{arocha3}. Since trellis based decoding of Nordstrom-Robinson code is too complex, a new SISO decoding was proposed \cite{ali1,axie1}. The lifting decoder technique for decoding of free linear codes over rings was introduced in \cite{agreferath1} and extended to arbitrary linear codes over rings in \cite{ababu1}. Chase decoding of linear $\mathbb{Z}_4$ codes based on the lifting decoder technique was presented in \cite{aarmond1,aarmond2}. Two algorithms for decoding of RM-like codes over rings of characteristic 2, based on coset decomposition were proposed in \cite{adavis1}. Though these algorithms primarily use hard-decision decoding, there is an extension to the soft-input case as well. Sphere decoding of Kerdock and Delsarte-Goethals codes was presented in \cite{acalderbank1}. Several algebraic decoders of codes over rings were presented in \cite{ahammons1,ahelleseth1,abyrne1,arong1,aranto1}.

\mypar{Contribution.} In this paper, we present symbol-wise MAP decoding algorithms for Kerdock and Preparata codes with complexity $\mathcal{O}(N^2\log_2 N)$ (where $N$ is the code length in $\mathbb{Z}_4$). This is the same order of complexity as the ML decoding algorithms of Kerdock codes, presented in \cite{ahammons1,aadoul1,aelia1}. To the best of our knowledge, no MAP decoding algorithms were presented for these codes. We further present a sub-optimal bitwise APP decoding algorithm based on the lifting decoder technique \cite{agreferath1} that performs within $1.5dB$ of the optimum and has the complexity of $\mathcal{O}(N \log_2 N)$. This sub-optimal decoder can be applied to any linear $\mathbb{Z}_4$ code, but the decoding complexity depends on the encoder and decoder complexity of its associated binary code. Although there are several soft input decoders, these are the first probabilistic SISO decoders for these code families (with the exception of the BCJR decoder, whose complexity was noted to be too high, and the Nordstrom-Robinson code) that we know of. We note that a low-complexity bit-wise SISO decoder for the Nordstrom-Robinson code, based on the binary nonlinear representation, was investigated in \cite{ali1,axie1}. As the quaternary Nordstrom-Robinson code belongs to both the Kerdock and the Preparata families, it can efficiently be decoded using any of the novel MAP decoders or the lifting decoder presented in this paper. This is demonstrated in Section \ref{sc_sim}. The novel decoding algorithms are compared to the original lifting decoder \cite{agreferath1} and the chase decoder \cite{aarmond1,aarmond2}, and it is shown that our decoders significantly outperform these algorithms for the case of both Kerdock and Preparata codes.

\mypar{Paper organization.} The remainder of this paper is composed of five sections. Section \ref{sc_model} introduces the necessary definitions and presents the system model analyzed in this paper. Section \ref{sc_map} introduces MAP decoding of Kerdock and Preparata codes. Section \ref{sc_app} presents a fast, sub-optimal, soft input, soft output decoding algorithm for Kerdock and Preparata codes. Section \ref{sc_sim} presents simulation results, and Section \ref{sc_fin} concludes the paper. 

\mypar{Notation.} Throughout this paper, bold letters are used to represent module elements\footnote{As every vector space is by definition a free module, we will also refer to vector spaces as modules and to vectors as elements.} and the standard function notation is used to represent polynomials. The $i$-th component (term) of an element $\boldsymbol{x}$ (polynomial $x(Z)$, where $Z$ is a dummy variable), is denoted $x_i$. Given two elements, $\boldsymbol{x}$ and $\boldsymbol{y}$ of length $N$, $\boldsymbol{x} \odot \boldsymbol{y} = [x_0y_0, x_1y_1, \dots, x_{N-1}y_{N-1}]$ represents the Hadamard (point-wise) product, and $\langle\boldsymbol{x};\boldsymbol{y}\rangle$ represents the scalar product. Upper case letters represent matrices, where $\boldsymbol{x}_n$ denotes the $n$-th row of some matrix $X$. $P[x]$ represents the probability of $x$ and $P[x|y]$ represents the conditional probability of $x$ given $y$. For convenience, $\delta_{a, b}$ represents the Kronecker delta, defined as $\delta_{a, b} = 1$ if $a = b$ and $
\delta_{a, b} = 0$ if $a \neq b$.  Standard blackboard bold letters are used for sets of numbers (and the corresponding rings/fields over them), i.e. $\mathbb{Z}$ is the set of integers, $\mathbb{R}$ is the set of real numbers and $\mathbb{C}$ is the set of complex numbers. $\mathbb{Z}_n = \mathbb{Z}/n\mathbb{Z}$ represents the set of integers modulo $n$. $\mathbb{K}^N$ represents the set of all $N$-tuples of a set $\mathbb{K}$. Cursive uppercase letters represent codes and other sets. Other notation is introduced as it is used. 

\section{System Model}
\label{sc_model}

Let operators $+$ and $\cdot$ represent addition and multiplication in $\mathbb{Z}_4$, while $\oplus$ and $\otimes$ represent addition and multiplication in $\mathbb{Z}_2$. Note that $\mathbb{Z}_2 \subset \mathbb{Z}_4$, so the elements $\{0, 1\} \in \mathbb{Z}_2$ can be regarded as the same elements in $\mathbb{Z}_4$. Furthermore, any element $\alpha \in \mathbb{Z}_4$ has a dyadic expansion (representation)
\begin{equation}
\label{dyadic_exp}
\alpha = a + 2\cdot b, \text{with } a, b \in \mathbb{Z}_2.
\end{equation}
We will often write the dyadic expansion of $\alpha$ as a tuple $(a,b)$, where identity $2a = 2\alpha$ follows from (\ref{dyadic_exp}). The dyadic expansion can naturally be extended to module elements and matrices over $\mathbb{Z}_4$. For completeness, we reproduce the \textit{Lemma 1} from \cite{agreferath1} for the case of $\mathbb{Z}_4 = \mathbb{Z}_{2^2}$, \myparni{\cite[Lemma 1]{agreferath1}:} Let $m, n$ be natural numbers and let $\boldsymbol{v} \in \mathbb{Z}_2^m$, $\boldsymbol{w} \in \mathbb{Z}_2^n$ and $G \in \mathbb{Z}_2^{m \times n}$. Then
\[
2 \cdot (\boldsymbol{v}\otimes G \oplus \boldsymbol{w}) = 2 \cdot (\boldsymbol{v} \cdot G + \boldsymbol{w}).
\]
This identity will be used later on.

An $[N,K = k_1 + k_2]$ linear code of type $4^{k_1}2^{k_2}$ over $\mathbb{Z}_4$ is a submodule of $\mathbb{Z}_4^N$, defined by a generator matrix $G$. If $k_2 = 0$, the corresponding code is a free submodule and the generator matrix can be written in the systematic form - consisting of the identity and the parity part. Both Kerdock and Preparata codes can be regarded as free submodules of $\mathbb{Z}_4^N$ \cite{ahammons1}. 

\subsection{Short Review of Kerdock and Preparata Codes}

Nonlinear binary Kerdock code of length $2^m$ (for some even $m$, $m \geq 4$) is constructed as a union of the RM$(1,m)$ code and its cosets in the RM$(2,m)$ \cite{macwilliams1977theory}, i.e. every Kerdock code is a supercode of the RM$(1,m)$ code and a subcode of the RM$(2,m)$ code \cite{akerdock1}. Similarly, the nonlinear binary Preparata code of length $2^m$ (for some even $m$, $m \geq 4$) can be constructed as a union of the RM$(m-3,m)$ code and its cosets in the RM$(m-2,m)$ \cite{macwilliams1977theory, apreparata1}. In the case of $m = 4$, we have the Nordstrom-Robinson code. It was shown in \cite{ahammons1} that these nonlinear codes can be represented as the binary images under the Gray map of cyclic codes over the $\mathbb{Z}_4$ ring.

When considering cyclic codes over fields it is common to talk about extensions of base fields. A similar approach is used for rings \cite{ahammons1}. We define a Galois ring as an extension of the base ring $\mathbb{Z}_4$. Let $h(Z)$ be an irreducible monic polynomial of order $m$ over $\mathbb{Z}_4$ (where $Z$ is the dummy variable), then the Galois ring $GR$ is isomorphic to the set of all polynomials modulo $h(Z)$, $\mathbb{Z}_4[Z]/h(Z)$. Every $GR$ has a primitive element $\xi$ of order $N-1$, where $N = 2^m$. It is clear that $\xi$ does not generate the entire ring, but every element $\alpha \in GR$ has a dyadic representation \cite{ahammons1}
\begin{equation}
\alpha = \xi^r + 2\xi^s, \ r, s \in \Delta = \{-\infty, 0, 1, \dots, N-2\},
\end{equation}
with a standard convention, $\xi^{-\infty} = 0$, and $\xi^{-\infty}\xi^r = 0$, $\forall r \in \Delta$.

The generator matrix of the extended Kerdock code $\mathcal{K}[2^{m}, m+1]$, is defined as \cite{ahammons1},
\begin{equation}
\label{eq:kergen}
G_\mathcal{K} = \begin{bmatrix}
1 & 1 & 1 & 1 &\cdots & 1 \\
0 & 1 & \xi & \xi^2 & \cdots & \xi^{N-2}  
\end{bmatrix} P_\pi,
\end{equation}
where $P_\pi$ is an $N \times N$ column permutation matrix. Let $(G_{\mathcal{K},0}, G_{\mathcal{K},1})$ represent the dyadic expansion of the generator matrix $G_\mathcal{K}$. The binary linear code defined by the generator matrix $G_{\mathcal{K},0}$ is called the associated binary code \cite{wan1997quaternary}. Matrix $G_{\mathcal{K},0}$ can be obtained from $G_\mathcal{K}$ by replacing $\xi^n, n \in \Delta$ with $\overline{\xi}^n, n \in \Delta$, where $\overline{\xi}$ is a primitive element of the irreducible monic polynomial $\overline{h}(Z) \in \mathbb{Z}_2[Z]$, chosen so that ${\overline{h}(Z) \equiv h(Z)\;(\text{mod }2)}$ \cite{ahammons1, wan1997quaternary}. Since $\overline{\xi}$ is a primitive root of order $N-1$, the set $\{\overline{\xi}^n \; | \; n \in \Delta\}$ corresponds to the set of all $m$-dimensional module elements over $\mathbb{Z}_2$, which implies that $G_{\mathcal{K},0}$ is in fact the generator matrix of the RM($1,m$) code. For a detailed proof see \cite[Corollary 8.4]{wan1997quaternary}. Column permutation matrix $P_\pi$ is chosen so that the associated binary code of the Kerdock code is the first order Reed-Muller code RM($1, m$) obtained from a Sylvester-type binary Hadamard matrix \cite{macwilliams1977theory}. This makes it possible to apply the fast Walsh-Hadamard transform, and significantly reduce the complexity of the decoding algorithm.

The extended Preparata code $\mathcal{P}[2^{m},2^m - m - 1]$, is the dual of the extended Kerdock code, defined by the parity check matrix $H_\mathcal{P} = G_\mathcal{K}$. We assume that the generator matrix $G_\mathcal{P}$ is given in a systematic form\footnote{By \textit{systematic form}, we assume that a subset of columns of the generator matrix form the identity matrix, while the remaining columns represent parity.} (this follows from the fact that $G_\mathcal{K}$ can easily be reduced to a systematic form by only row operations). It can easily be verified that the associated binary code of the Preparata code is in fact the RM($m-2,m$) code \cite[Corollary 9.3]{wan1997quaternary}. Note that the encoding complexity of both the Kerdock and Preparata code is $\mathcal{O}(N\log_2 N)$. This is an important fact that will be used later on.

%

\subsection{Communication System and MAP Decoding}
\label{ss:mapmodel}

Let $\mathcal{C}$ be an arbitrary $[N, K]$ linear block code over some ring $\mathbb{K}$, and let all the codewords be equally likely. Given a codeword $\boldsymbol{c} \in \mathcal{C}$ and a modulation mapping $\phi(\cdot)$, the corresponding modulated codeword is $\boldsymbol{x} = \phi(\boldsymbol{c})$. The random mapping $\Omega(\cdot)$ represent a memoryless channel, defined by a conditional probability 
\begin{equation}
P[\boldsymbol{y}|\boldsymbol{c}] = \prod_{n=0}^{N-1} P[y_n | c_n],
\end{equation}
where $\boldsymbol{y} = \Omega(\boldsymbol{x})$ is a specific channel output. For every codeword symbol $c_j$, $j \in \{0, 1, \dots N-1\}$, the goal of the MAP decoder is to find ${P[c_j = \alpha | \boldsymbol{y}]}$ for every $\alpha \in \mathbb{K}$ and choose the maximum one. 
Using Bayes's rule and the definition of the memoryless channel, we can write
\begin{equation}
\label{eq:map_deco_rule}
\begin{split}
P[c_j = \alpha|\boldsymbol{y}] &= \sum_{\boldsymbol{b} \in \mathcal{C}}P[\boldsymbol{b} | \boldsymbol{y}] \delta_{b_j, \alpha} \\ &=  \sum_{\boldsymbol{b} \in \mathcal{C}} \frac{P[\boldsymbol{b}]}{P[\boldsymbol{y}]}  P[\boldsymbol{y} | \boldsymbol{b}] \delta_{b_j, \alpha} \\ &= \frac{1}{\vert \mathcal{C} \vert P[\boldsymbol{y}]} \sum_{ \boldsymbol{b} \in \mathcal{C}} \prod_{n = 0}^{N-1} P[y_n | b_n] \delta_{b_j, \alpha},
\end{split}
\end{equation}
where $|\mathcal{C}|$ represents the size of the code $\mathcal{C}$.

A MAP decoding rule for codes over finite fields using the dual code was introduced in \cite{ahartman1} and extended to codes over rings in \cite{aberkamn1, aberkamn2}. 
Let the Fourier transform ($\mathscr{F}\{\cdot\}$) of a real or complex-valued function $f(\alpha)$, $\alpha \in \mathbb{K}$, be defined as
\begin{equation}
\label{eq:ft_def}
\mathscr{F}\{ f(\alpha) \} = \sum_{\alpha \in \mathbb{K}} \omega^{-\alpha\beta} f(\alpha)  = F(\beta), \ \beta \in \mathbb{K},
\end{equation}
with 
\begin{equation}
\omega = \exp \left\{\frac{2\pi I}{|\mathbb{K}|}\right\}, \ I=\sqrt{-1}.
\end{equation}

Using the Poisson summation formula and the fact that the Fourier transform of a product function is the product of the corresponding individual Fourier transforms \cite{aberkamn2, aforney2}, the MAP decoding rule in (\ref{eq:map_deco_rule}), can be written as
\[
P[c_j = \alpha | \boldsymbol{y}] = \\ \frac{1}{|\mathbb{K}|^N P[\boldsymbol{y}]} \sum_{\boldsymbol{b} \in \mathcal{C}^\bot} \prod_{n=0}^{N-1} \mathscr{F}\{P[y_n|b_n]\} \mathscr{F}\{\delta_{b_j,\alpha}\},
\]
where $\mathcal{C}^\bot$ is the dual code of $\mathcal{C}$ and $\vert \mathbb{K} \vert^N = \vert \mathcal{C} \vert \cdot \vert \mathcal{C}^\bot \vert$. By substituting eq. (\ref{eq:ft_def}) for the Kronecker delta we get
\begin{equation}
\label{eq:dualsub}
\begin{split}
P[c_j = \alpha | \boldsymbol{y}] &= \eta \sum_{\boldsymbol{b} \in \mathcal{C}^\bot}\prod_{n=0}^{N-1} \mathscr{F}\{P[y_n|b_n]\} \sum_{\beta \in \mathbb{K}} \omega^{-\alpha\beta} \delta_{b_j, \beta} \\ &=
\eta  \sum_{\beta \in \mathbb{K}} \omega^{-\alpha\beta} \sum_{\boldsymbol{b} \in \mathcal{C}^\bot} \prod_{n=0}^{N-1} \mathscr{F}\{P[y_n|b_n]\} \delta_{b_j, \beta},
\end{split}
\end{equation} 
where $\eta = \frac{1}{|\mathbb{K}|^N P[\boldsymbol{y}]}$. After substituting eq. (\ref{eq:ft_def}) in (\ref{eq:dualsub}) we have
\begin{multline}
\label{dualsub2}
P[c_j = \alpha | \boldsymbol{y}] = \\ \eta  \sum_{\beta \in \mathbb{K}} \omega^{-\alpha\beta} \sum_{\boldsymbol{b} \in \mathcal{C}^\bot} \left[\prod_{n=0}^{N-1} \sum_{\gamma \in \mathbb{K}} \omega^{-\gamma b_n} P[y_n | \gamma] \cdot \delta_{b_j, \beta} \right].
\end{multline}
By using the identity $\delta_{a,b} = \frac{1}{|\mathbb{K}|} \sum_{\gamma \in \mathbb{K}} \omega^{\gamma \cdot (a - b)}$ and a little arithmetic, we get
\begin{equation}
\label{eq:dualmap}
\begin{split}
P[c_j = \alpha | \boldsymbol{y}] = \frac{\eta}{|\mathbb{K}|}  A_j(\alpha), \ \alpha \in \mathbb{K}
\end{split}
\end{equation}
where

\[
A_j(\alpha) = \sum_{\beta \in \mathbb{K}} \omega^{-\alpha\beta} \sum_{\boldsymbol{b} \in \mathcal{C}^\bot} \left[\prod_{n=0}^{N-1} \sum_{\gamma \in \mathbb{K}} \omega^{-\gamma(b_n - \beta \delta_{j,n})} P[y_n | \gamma]\right].
\]

\section{MAP Decoding Algorithm}
\label{sc_map}

Let $\mathbb{S} = \mathbb{R}[Z]/\langle Z^4 - 1\rangle$, be a set of all polynomials over $\mathbb{R}$, modulo $Z^4 - 1$, where $Z$ is a dummy variable. Note that the multiplicative group of monic monomials in $\mathbb{S}$ is equivalent to the additive group of $\mathbb{Z}_4$. For convenience, we define a mapping $\zeta : \mathbb{Z}_4 \rightarrow \mathbb{S}$, such that $\zeta(\alpha) = Z^\alpha$, where a natural extension to elements and matrices is to apply $\zeta(\cdot)$ componentwise. Given $f(Z) = \sum_{\alpha \in \mathbb{Z}_4} f_\alpha Z^\alpha \in \mathbb{S}$ and $Z^{-\beta} \in \mathbb{S}$, $\beta \in \mathbb{Z}_4$, then

\begin{equation}
f(Z) \cdot Z^{-\beta} = \sum_{\alpha \in \mathbb{Z}_4} f_{\alpha} Z^{\alpha - \beta} = \sum_{\alpha \in \mathbb{Z}_4} f_{\alpha + \beta} Z^\alpha \in \mathbb{S}.
\end{equation}

Furthermore, let $\mathcal{C}' \subseteq \mathcal{C}$ be a linear $[N,K-1]$ subcode of $\mathcal{C}$, such that $\boldsymbol{1} \notin \mathcal{C}'$, where $\boldsymbol{1}$ is the all-one element of length $N$. We assume that all codewords $\boldsymbol{c} \in \mathcal{C}'$ are enumerated, and arranged in a matrix $A$, such that the $k$-th row of $A$ is equal to the $k$-th codeword $\boldsymbol{c}_k$. Similar to denotations in \citep{aashikhmin1}, let $D = \zeta(A)$ and $\overline{D} = \zeta(-A)$, be two $4^{K-1} \times N$ matrices.

We again assume a codeword $\boldsymbol{c} \in \mathcal{C}$ is transmitted over a memoryless channel, and $\boldsymbol{y}$ is received. Following the group algebra description of MAP decoding from \cite{aashikhmin1}, we first compute the reliability element $\boldsymbol{w} \in \mathbb{S}^N$, where each component is a log-likelihood polynomial, defined as
\begin{equation}
\label{eq:likelihoods}
w_n(Z) = \sum_{\alpha \in \mathbb{Z}_4} \log P[y_n | \alpha] Z^\alpha,\ n = 0, \dots, N-1.
\end{equation}
We compute sums of logarithms of probabilities of receiving the channel output $\boldsymbol{y}$ conditioned on the sent codeword, as
\begin{equation}
\label{eq:hardway}
\boldsymbol{t}^\top = \overline{D} \boldsymbol{w}^\top \in \mathbb{S}^{4^{K-1}},
\end{equation}
where we use \textit{Proposition 2} from \cite{aashikhmin1}, which we summarize here, \myparni{\cite[Proposition 2]{aashikhmin1}:} Given a reliability element $\boldsymbol{w}$ and a codeword $\boldsymbol{c}_k \in \mathcal{C}$ the following identity holds:
\begin{equation}
\begin{split}
t_k(Z) &= \langle \boldsymbol{w} ; \zeta(-\boldsymbol{c}_k)\rangle \\ &= \sum_{\alpha \in \mathbb{Z}_4} t_{k,\alpha} Z^\alpha = \sum_{\alpha \in \mathbb{Z}_4} \sum_{n = 0}^{N-1} \log P[y_n|\alpha + c_{k,n}] Z^{\alpha} \\ &= 
\sum_{\alpha \in \mathbb{Z}_4} \log P[\boldsymbol{y}|\boldsymbol{c}_k + \alpha\boldsymbol{1}] Z^\alpha, \boldsymbol{c}_k \in \mathcal{C}.
\end{split}
\end{equation}

Next, we recompute the sums of logarithms of probabilities into products of probabilities, $\boldsymbol{v} \in \mathbb{S}^{4^{K-1}}$, as follows
\begin{equation}
v_k(Z) = \sum_{\alpha \in \mathbb{Z}_4} \text{e}^{t_{k,\alpha}}Z^\alpha, \ k = 0, \dots, 4^{K-1}-1.
\end{equation}
Finally, we compute soft decisions $\boldsymbol{s}^\top = D^\top \boldsymbol{v}^\top \in \mathbb{S}^N$, where
\begin{equation}
s_n(Z) = \sum_{\alpha \in \mathbb{Z}_4} Z^\alpha \sum_{\boldsymbol{b} \in \mathcal{C}} P[\boldsymbol{y} | \boldsymbol{b}] \delta_{b_j,\alpha}, \ n = 0, \dots, N-1.
\end{equation}

The MAP decision is by definition, $c_n = \beta$, such that $s_{n,\beta} = \max_{\alpha \in \mathbb{Z}_4} \{s_{n,\alpha}\}$ (i.e. $\beta$ is the index of the largest real-valued term in the polynomial $s_n(Z) \in \mathbb{S}$), for all $n = 0, \dots, N-1$. It is clear that the most computationally expansive steps are multiplications with $D$ and $\overline{D}$. In the following subsection, we will show how this can be simplified in the case of the Kerdock code. In the case of the Preparata code, we will use the dual MAP decoding rule, given in (\ref{eq:dualmap}). 

Consider the codeword $\boldsymbol{c} = \boldsymbol{u} \cdot G \in \mathbb{Z}_4^N$, where $G \in \mathbb{Z}_4^{K \times N}$ is the generator matrix and $\boldsymbol{u} \in \mathbb{Z}_4^K$ is the corresponding information sequence. Let $(\boldsymbol{u}_0, \boldsymbol{u}_1)$ and $(G_0, G_1)$ represent the dyadic expansions of the information sequence and the generator matrix, respectively. From the dyadic expansion of the information sequence, it follows that
\begin{equation}
\label{eq:half_exp}
\boldsymbol{c} = \boldsymbol{u} \cdot G = (\boldsymbol{u}_0 + 2\cdot\boldsymbol{u}_1) \cdot G = \boldsymbol{u}_0 \cdot G + 2\cdot(\boldsymbol{u}_1 \otimes G_0),
\end{equation}
where the last equality follows from \textit{Lemma 1} in \cite{agreferath1}. It is well known that in the case of the Kerdock and Preparata codes, the $G_0$ is the generator matrix of the corresponding RM subcode \cite{ahammons1}. 

\subsection{MAP Decoding of Kerdock codes}

Let $G$ be the Kerdock generator matrix, defined in (\ref{eq:kergen}). We will define three auxiliary matrices, $A_0 \in \mathbb{Z}^{4 \times N}$, $A_1 \in \mathbb{Z}^{N \times N}$ and $A_2 \in \mathbb{Z}^{N \times N}$, together with their row sets (sets of row elements), $\mathcal{A}_0$, $\mathcal{A}_1$ and $\mathcal{A}_2$. Let the rows of $A_0$ be $\alpha\boldsymbol{1}$, $\alpha \in \mathbb{Z}_4$. Let the rows of $A_1$ be equal to $\boldsymbol{i} \cdot G$, where $\boldsymbol{i} \in \mathbb{Z}_2^K$ is a binary sequence of length $K$, such that the first value is always zero. Finally, let the rows of $A_2$ be equal to $2\cdot(\boldsymbol{i} \otimes G_0)$, where $\boldsymbol{i}$ is the same as in the previous case. Note that $A_2 = 2 \cdot A_1 = 2 \cdot H_{2^m}$, where $H_{2^m}$ is the binary Hadamard matrix, which follows from the fact that the associated binary code of the Kerdock code is the RM($1,m$) code. We can now define the Kerdock code as
\begin{equation}
\mathcal{K} = \{\boldsymbol{a}^0 + \boldsymbol{a}^1 + \boldsymbol{a}^2 | \boldsymbol{a}^0 \in \mathcal{A}_0, \boldsymbol{a}^1 \in \mathcal{A}_1, \boldsymbol{a}^2 \in \mathcal{A}_2 \},
\end{equation}
while the linear subcode $\mathcal{K}' \subseteq \mathcal{K}$, that does not contain the all-one codeword, can be represented as
\begin{equation}
\mathcal{K}' = \{\boldsymbol{a}^1 + \boldsymbol{a}^2 | \boldsymbol{a}^1 \in \mathcal{A}_1, \boldsymbol{a}^2 \in \mathcal{A}_2 \}.
\end{equation}
Note that this representation is equivalent to the trace representation of the Kerdock code, introduced in \cite{ahammons1}. Let $D_1 = \zeta(A_1)$ and $D_2 = \zeta(A_2)$. Similarly, let $\overline{D}_1 = \zeta(-A_1)$ and $\overline{D}_2 = \zeta(-A_2) = D_2$. As any row of $A$ can be represented as a sum of rows from $A_1$ and $A_2$, it follows that any row $\boldsymbol{d}_k = \zeta(\boldsymbol{c}_k)$ of $D$ can be represented as $\boldsymbol{d}_n^1 \odot \boldsymbol{d}_l^2$, where $\boldsymbol{d}_n^1$ is a row from $D_1$ and $\boldsymbol{d}_l^2$ is a row from $D_2$.

Given the channel output $\boldsymbol{y}$, we again start the decoding process by computing $\boldsymbol{w} \in \mathbb{S}^N$ using eq. (\ref{eq:likelihoods}). We note that 
\begin{equation}
t_k(Z) = \langle \boldsymbol{w} ; \overline{\boldsymbol{d}}_n^1 \odot \boldsymbol{d}_l^2 \rangle = \langle \boldsymbol{w} \odot \overline{\boldsymbol{d}}_n^1; \boldsymbol{d}_l^2 \rangle,
\end{equation}
where the last equality follows from the commutativity of multiplication. Instead of applying eq. (\ref{eq:hardway}), we first define a matrix $B \in \mathbb{S}^{N \times N}$, with rows $\boldsymbol{b}_n = \boldsymbol{w} \odot \overline{\boldsymbol{d}}^1_n$, $n = 0, \dots, N-1$. Note that every component of $B$ is equal to
\begin{equation}
b_{n,l}(Z) = \sum_{\alpha \in \mathbb{Z}_4} \log P[\boldsymbol{y_l} | a^1_{n,l} + \alpha ] Z^\alpha, \ n, l = 0, \dots, N-1.
\end{equation}
Next, we will compute the matrix of the sums of logarithms of probabilities
\begin{equation}
T = B \cdot D_2^{\text{T}}.
\end{equation}
This multiplication can be efficiently implemented by applying a modified fast Walsh-Hadamard transform (FWHT), presented in Algorithm \ref{alg1fwht}, to every row of $B$, $\boldsymbol{t}_n = \text{FWHT} (\boldsymbol{b}_n)$, $n = 0, \dots N-1$. Note that every component of $T$ is equal to
\begin{equation}
\begin{split}
t_{n,l}(Z) = \sum_{\alpha \in \mathbb{Z}_4} Z^\alpha \sum_{j = 0}^{N-1} \log P[\boldsymbol{y_l} | a^1_{n,j} + a^2_{l,j} + \alpha ], \\ \ n,l = 0, \dots, N-1.
\end{split}
\end{equation}

\begin{figure}[bt]
\begin{algorithm}[H]
\centering
\begin{algorithmic}[1]
 \renewcommand{\algorithmicrequire}{\textbf{Input:}}
 \renewcommand{\algorithmicensure}{\textbf{Output:}}
\REQUIRE $\boldsymbol{a}$,
\ENSURE $\boldsymbol{a}$,
\FOR{$h = 1$; $h < N$; $h \leftarrow 2\cdot h$}
\FOR{$i = 0$; $i < N$; $i \leftarrow i + 2\cdot h$}
\FOR{$j = i$ \TO $i + h$}
\STATE $x \leftarrow a_j$, $y \leftarrow a_{j+h}$,  \COMMENT{Save current values}
\STATE $a_j \leftarrow x + y$, \COMMENT{Calculate new $a_j$ value}
\STATE $a_{j+h} \leftarrow x + Z^2 \cdot y$, \COMMENT{Calculate new $a_{j+h}$ value}
\ENDFOR
\ENDFOR
\ENDFOR
\end{algorithmic}
\caption{In-place Fast Walsh-Hadamard transform}
\label{alg1fwht}
\end{algorithm}
\end{figure}

Let $V \in \mathbb{S}^{N\times N}$ be a matrix of products of probabilities, with components
\begin{equation}
\begin{split}
v_{n,l}(Z) &= \sum_{\alpha \in \mathbb{Z}_4} Z^\alpha \exp\left\{t_{n,l,\alpha}\right\} \\ &= \sum_{\alpha \in \mathbb{Z}_4} Z^\alpha \prod_{j=0}^{N-1} P[\boldsymbol{y_l} | a^1_{n,j} + a^2_{l,j} + \alpha ] \\ &= \sum_{\alpha \in \mathbb{Z}_4} P[\boldsymbol{y_l} | \boldsymbol{c}_k + \alpha\boldsymbol{1}] Z^\alpha,
\end{split}
\end{equation}
where $\boldsymbol{c}_k = \boldsymbol{a}^1_n + \boldsymbol{a}^2_l$ and $n,l = 0, \dots, N-1$. 

Next, we compute $Q = V \cdot D_2 \in \mathbb{S}^{N \times N}$, with
\begin{equation}
\begin{split}
q_{n,l}(Z) &= \sum_{j = 0}^{N-1} Z^{a^2_{l,j}} \cdot v_{n,j} \\ &= \sum_{\alpha \in \mathbb{Z}_4} Z^\alpha \sum_{j = 0}^{N-1} P[\boldsymbol{y}|\boldsymbol{c}_k + (\alpha - a^2_{l,j})\boldsymbol{1}].
\end{split}
\end{equation}
where $\boldsymbol{c}_k = \boldsymbol{a}^1_n + \boldsymbol{a}^2_j$, and $n,l=0, \dots, N-1$. This product can also be efficiently implemented by applying FWHT transforms to every row of $V$. Finally, let the element of soft decisions $\boldsymbol{s} \in \mathbb{S}^N$ be equal to 
\begin{equation}
\boldsymbol{s} = \sum_{j=0}^{N-1} \boldsymbol{d}_j^1 \odot \boldsymbol{q}_j,
\end{equation}
with
\begin{equation}
\begin{split}
s_n(Z) &= \sum_{i = 0}^{N-1} Z^{a^2_{i,n}} \cdot q_{n,l} \\ &= \sum_{\alpha \in \mathbb{Z}_4} \sum_{j = 0}^{N-1} \sum_{i = 0}^{N-1} P[\boldsymbol{y} | \boldsymbol{c}_k + (\alpha - a^2_{l,j} - a^1_{i,n})\boldsymbol{1}] \\ &= \sum_{\alpha \in \mathbb{Z}_4} Z^\alpha \sum_{\boldsymbol{b} \in \mathcal{C}} P[\boldsymbol{y} | \boldsymbol{b}] \delta_{b_j,\alpha}, \ n = 0, \dots, N-1.
\end{split}
\end{equation}
This concludes the algorithm. We summarize the MAP decoding of Kerdock codes in Algorithm \ref{alg2mapk}.

\begin{figure}[bt]
\begin{algorithm}[H]
\centering
\begin{algorithmic}[1]
 \renewcommand{\algorithmicrequire}{\textbf{Input:}}
 \renewcommand{\algorithmicensure}{\textbf{Output:}}
\REQUIRE $\boldsymbol{y}$,
\ENSURE $\boldsymbol{s}$
\STATE Compute the reliability element $\boldsymbol{w}$ using eq. (\ref{eq:likelihoods}).
\FOR{$n = 0$; $n < N$; $n \leftarrow n + 1$}
\STATE $\boldsymbol{t}_n \leftarrow \text{FWHT}(\boldsymbol{w} \odot \overline{\boldsymbol{d}}^1_n)$,  
\ENDFOR
\FOR{$n = 0$; $n < N$; $n \leftarrow n + 1$}
\STATE \COMMENT{Compute element $\boldsymbol{v}_n \in \mathbb{S}^{N}$ as follows}
\FOR{$l = 0$; $l < N$; $l \leftarrow l + 1$}
\STATE $v_{n,l}(Z) \leftarrow \sum_{\alpha \in \mathbb{Z}^4} Z^\alpha \exp \{t_{n,l,\alpha}\}$
\ENDFOR
\ENDFOR
\STATE $\boldsymbol{s} \leftarrow \boldsymbol{0}$ \COMMENT{Initialize $\boldsymbol{s}$ to the all-zero value}
\FOR{$n = 0$; $n < N$; $n \leftarrow n + 1$}
\STATE $\boldsymbol{s} \leftarrow \boldsymbol{s} + \text{FWHT}(\boldsymbol{v}_n) \odot \boldsymbol{d}^1_n$,  
\ENDFOR
\end{algorithmic}
\caption{MAP decoding of Kerdock codes}
\label{alg2mapk}
\end{algorithm}
\end{figure}

Calculating $\boldsymbol{s}$ and forming matrices $B$ and $V$ have complexity $\mathcal{O}(N^2)$, while for computing the matrices $T$ and $Q$, we have $N$ uses of the FWHT transform, with complexity $\mathcal{O}(N\log_2 N)$, so the total complexity of this decoding algorithm is $\mathcal{O}(N^2 \log_2 N)$. A detailed complexity analysis of Algorithm \ref{alg2mapk} is given in Table \ref{minja1t}. All operations are assumed to be carried out over the polynomial ring $\mathbb{S}$. As polynomials in $\mathbb{S}$ can efficiently be implemented as floating-point arrays of length 4, with multiplication implemented as circular convolution (with complexity $4^2\mathcal{O}(1)$) and addition implemented as component-wise addition (with complexity $4\mathcal{O}(1)$), the additional computational cost can be treated as a constant term. 

\begin{table}[!tb]
\caption{{The computational cost of MAP decoding of Kerdock codes}}
\label{minja1t}
\centering
\begin{tabular}{|c|c|c|} 
\hline
\parbox[t]{1.2cm}{\centering Steps} & {\centering Short description} & {\centering Complexity} \\ \hline \hline
1. & Compute $\boldsymbol{w}$ & $\mathcal{O}(N)$ \tabularnewline \hline 
2. - 4. & FWHT of $\boldsymbol{w} \odot \overline{\boldsymbol{d}}^1_n$ & $N \mathcal{O}(N\log_2 N) = \mathcal{O}(N^2\log_2 N)$ \tabularnewline \hline
5. - 10. & Calculate matrix $V$ & $\mathcal{O}(N^2)$ \tabularnewline \hline 
11. & Initialize $\boldsymbol{s}$ & $\mathcal{O}(N)$ \tabularnewline \hline 
12. - 14. & Compute $Q$ and $\boldsymbol{s}$ & $N\mathcal{O}(N\log_2 N) = \mathcal{O}(N^2\log_2 N)$ \tabularnewline \hline 
\end{tabular}
\end{table}


\subsection{MAP Decoding of Preparata codes}
Consider a Preparata code with a parity-check matrix defined in eq. (\ref{eq:kergen}). Given the channel output $\boldsymbol{y}$, we begin the decoding process by calculating the element of likelihoods $\boldsymbol{w} \in \mathbb{S}^N$, with 
\begin{equation}
\label{eq:calclikelihoodp}
w_n(Z) = \sum_{\alpha \in \mathbb{Z}_4} P[y_n | \alpha] Z^\alpha.
\end{equation}
Next, we calculate the element $\boldsymbol{r}$ as the componentwise Fourier transform of $\boldsymbol{w}$, with
\begin{equation}
\label{eq:fftcwp}
r_n(Z) = \mathscr{F}(\boldsymbol{w}_n) = \sum_{\alpha \in \mathbb{Z}_4} Z^\alpha \sum_{\gamma \in \mathbb{Z}_4} \omega^{-\alpha\gamma} P[y_n | \gamma].
\end{equation}

Let functions $|\cdot|$ and $\arg \{\cdot\}$ represent the modulus and the argument of a complex number, respectively. We will form elements $\boldsymbol{\rho} \in \mathbb{S}^N$ and $\boldsymbol{\phi} \in \mathbb{S}^N$, with components
\begin{equation}
\label{eq:rhop}
\rho_n(Z) = \sum_{\alpha \in \mathbb{Z}_4} \log \vert r_{n,\alpha} \vert Z^\alpha,
\end{equation}
and
\begin{equation}
\label{eq:phip}
\phi_n(Z) = \sum_{\alpha \in \mathbb{Z}_4} \arg\{r_{n,\alpha}\} Z^\alpha.
\end{equation}

Next, we compute elements $\boldsymbol{t} \in \mathbb{S}^{4^{K-1}}$ and $\boldsymbol{e} \in \mathbb{S}^{4^{K-1}}$ as
\begin{equation}
\boldsymbol{t}^T = \overline{D} \boldsymbol{\rho}^T,
\end{equation}
and
\begin{equation}
\boldsymbol{e}^T = \overline{D} \boldsymbol{\phi}^T.
\end{equation}
Note that similarly as in the case of the Kerdock code, we can efficiently calculate the element $\boldsymbol{t}$, by first computing the matrix $B \in \mathbb{S}^{N \times N}$ with rows $\boldsymbol{b}_n = \boldsymbol{\rho} \odot \overline{\boldsymbol{d}}_n^1$, $n = 0, \dots, N-1$, and then computing the matrix $T \in \mathbb{S}^{N \times N}$ with rows $\boldsymbol{t}_n = \text{FWHT}(\boldsymbol{b}_n)$, $n = 0, \dots N-1$. Every component of $\boldsymbol{t}$ now corresponds to one component of matrix $T$. We apply the same procedure to get the matrix $E \in \mathbb{S}^{N \times N}$ that corresponds to element $\boldsymbol{e}$.

Next, we form an $N \times N$ matrix $V$ with components
\begin{equation}
\label{eq:matvp}
v_{n,l}(Z) = \sum_{\alpha \in \mathbb{Z}_4} \exp\{t_{n,l,\alpha}\} \cdot \exp\{e_{n,l,\alpha}\} Z^\alpha.
\end{equation}

Let $Q$ be an $N \times N$ matrix with rows $\boldsymbol{q}_n = \text{FWHT}(\boldsymbol{v}_n)$, $n = 0, \dots, N-1$. We calculate element $\boldsymbol{s}$ as 
\begin{equation}
\boldsymbol{s} = \sum_{n = 0}^{N-1} \boldsymbol{d}^1_n \odot \boldsymbol{q}_n,
\end{equation}
with
\begin{multline}
s_j(Z) = \\ \sum_{\alpha \in \mathbb{Z}_4} Z^\alpha \sum_{\boldsymbol{b} \in \mathcal{C}^\bot}  \delta_{b_j,\alpha} \prod_{n=0}^{N-1} \sum_{\gamma \in \mathbb{Z}_4} \omega^{-\gamma(\alpha - b_j + b_n)} P[y_n | \gamma].
\end{multline}
Note that this is equivalent as 
\begin{equation}
\boldsymbol{s} = D^T \boldsymbol{v}^T,
\end{equation}
where $\boldsymbol{v}$ is an element that corresponds to matrix $V$. Next, we calculate $\boldsymbol{g}$, as
\begin{equation}
\label{eq:calcqp}
g_{j}(Z)= \sum_{\alpha \in \mathbb{Z}_4} Z^\alpha \sum_{\beta \in \mathbb{Z}_4} \frac{s_{j,\beta}}{r_{j,\beta}} \cdot r_{j,\beta - \alpha}.
\end{equation}
After a simple analysis, we can see that 
\begin{multline}
g_j(Z) = \\ \sum_{\alpha \in \mathbb{Z}_4} Z^\alpha \sum_{\boldsymbol{b} \in \mathcal{C}^\bot} \prod_{n=0}^{N-1} \sum_{\gamma \in \mathbb{Z}_4} \omega^{-\gamma(b_n - b_j + \alpha \delta_{j,n})} P[y_n | \gamma].
\end{multline}

Next, with the help of (\ref{eq:dualmap}), we obtain,
\begin{equation}
\label{eq:calchp}
h_j(Z) =  \sum_{\alpha \in \mathbb{Z}_4} Z^\alpha \sum_{\beta \in \mathbb{Z}_4} \omega^{-\alpha\beta} g_{j,\beta}.
\end{equation}

Finally, we compute
\begin{equation}
\label{eq:calczp}
z_j(Z) = \sum_{\alpha \in \mathbb{Z}_4} Z^\alpha \frac{h_{j,\alpha}}{4\cdot m_j},
\end{equation}
where $m_j$ is a scaling factor defined as
\begin{equation}
\label{eq:calcmp}
m_{j} = \sum_{\beta \in \mathbb{Z}_4} \frac{s_{j,\beta}}{r_{j,\beta}}.
\end{equation}

This concludes the algorithm. It is easy to see that the complexity of this algorithm is the same as that of the Kerdock MAP decoding algorithm, i.e., $\mathcal{O}(N^2\log_2 N)$. We summarize the MAP decoding of Preparata codes in Algorithm \ref{alg2mapp}. A detailed complexity analysis of Algorithm \ref{alg2mapp} is given in Table \ref{minja2t}. Similar to the case of the Kerdock MAP decoder, all operations are done in $\mathbb{S}$. Note that the Fourier transform in eq. (\ref{eq:fftcwp}) also has complexity $4^2\mathcal{O}(1)$.

\begin{figure}[bt]
\begin{algorithm}[H]
\centering
\begin{algorithmic}[1]
 \renewcommand{\algorithmicrequire}{\textbf{Input:}}
 \renewcommand{\algorithmicensure}{\textbf{Output:}}
\REQUIRE $\boldsymbol{y}$,
\ENSURE $\boldsymbol{z}$
\STATE Compute the element of likelihoods $\boldsymbol{w}$ using eq. (\ref{eq:calclikelihoodp}).
\STATE Compute the element $\boldsymbol{r}$ using eq. (\ref{eq:fftcwp}).
\FOR{$n = 0$; $n < N$; $n \leftarrow n + 1$}
\STATE $\rho_n \leftarrow \sum_{\alpha \in \mathbb{Z}_4} \log \vert r_{n,\alpha} \vert Z^\alpha$,
\COMMENT{Using eq. (\ref{eq:rhop})}  
\STATE $\phi_n(Z) \leftarrow \sum_{\alpha \in \mathbb{Z}_4} \arg\{r_{n,\alpha}\} Z^\alpha$
\COMMENT{Using eq. (\ref{eq:phip})} 
\ENDFOR
\FOR{$n = 0$; $n < N$; $n \leftarrow n + 1$}
\STATE $\boldsymbol{t}_n \leftarrow \text{FWHT}(\boldsymbol{\rho}\odot\overline{\boldsymbol{d}}^1_n)$ 
\STATE $\boldsymbol{e}_n \leftarrow \text{FWHT}(\boldsymbol{\phi}\odot\overline{\boldsymbol{d}}^1_n)$ 
\ENDFOR
\FOR{$n = 0$; $n < N$; $n \leftarrow n + 1$}
\STATE \COMMENT{Compute element $\boldsymbol{v}_n \in \mathbb{S}^{N}$ using eq. (\ref{eq:matvp}):}
\FOR{$l = 0$; $l < N$; $l \leftarrow l + 1$}
\STATE $v_{n,l}(Z) \leftarrow \sum_{\alpha \in \mathbb{Z}^4} Z^\alpha \exp \{t_{n,l,\alpha}\}\cdot \exp \{e_{n,l,\alpha}\}$
\ENDFOR
\ENDFOR
\STATE $\boldsymbol{s} \leftarrow \boldsymbol{0}$ \COMMENT{Initialize $\boldsymbol{s}$ to the all-zero value}
\FOR{$n = 0$; $n < N$; $n \leftarrow n + 1$}
\STATE $\boldsymbol{s} \leftarrow \boldsymbol{s} + \text{FWHT}(\boldsymbol{v}_n) \odot \boldsymbol{d}^1_n$,  
\ENDFOR
\STATE Compute element $\boldsymbol{g}$ using eq. (\ref{eq:calcqp}).
\STATE Compute element $\boldsymbol{h}$ using $\boldsymbol{g}$ and eq. (\ref{eq:calchp}).
\FOR{$n = 0$; $n < N$; $n \leftarrow n + 1$}
\STATE $z_n(Z) \leftarrow \sum_{\alpha \in \mathbb{Z}_4} Z^\alpha \frac{h_{n,\alpha}}{4\cdot m_n},$,  
\COMMENT {Using eq. (\ref{eq:calczp}) and (\ref{eq:calcmp})}
\ENDFOR
\end{algorithmic}
\caption{MAP decoding of Preparata codes}
\label{alg2mapp}
\end{algorithm}
\end{figure}

\begin{table}[!tb]
\caption{{The computational cost of MAP decoding of Preparata codes}}
\label{minja2t}
\centering
\begin{tabular}{|c|c|c|} 
\hline
\parbox[t]{1.2cm}{\centering Steps} & {\centering Short description} & {\centering Complexity} \\ \hline \hline
1. & Compute $\boldsymbol{w}$ & $\mathcal{O}(N)$ \tabularnewline \hline 
2. & Compute $\boldsymbol{r}$ & $\mathcal{O}(N)$ \tabularnewline \hline 
3. - 6. & Compute $\boldsymbol{\rho}$ and $\boldsymbol{\phi}$ & $\mathcal{O}(N)$ \tabularnewline \hline 
7. - 10. & Compute $T$ and $E$ & $N \mathcal{O}(N \log_2 N) = \mathcal{O}(N^2\log_2 N)$ \tabularnewline \hline
11. - 16. & Calculate matrix $V$ & $\mathcal{O}(N^2)$ \tabularnewline \hline 
17. & Initialize $\boldsymbol{s}$ & $\mathcal{O}(N)$ \tabularnewline \hline 
18. - 20. & Compute $V$ and $\boldsymbol{s}$ & $N\mathcal{O}(N\log_2 N) = \mathcal{O}(N^2\log_2 N)$ \tabularnewline \hline 
21. & Compute $\boldsymbol{g}$ & $\mathcal{O}(N)$ \tabularnewline \hline 
22. & Compute $\boldsymbol{h}$ & $\mathcal{O}(N)$ \tabularnewline \hline 
23. - 25. & Compute $\boldsymbol{z}$ & $\mathcal{O}(N)$ \tabularnewline \hline 
\end{tabular}
\end{table}

\section{Bitwise APP Decoding Algorithm}
\label{sc_app}

A linear block code over $\mathbb{Z}_{p^k}$ (for some prime $p$ and a positive integer $k$) with a p-adic expansion of the generator matrix $(G_0, G_1, \dots G_{k-1})$, such that $G_0$ is of full rank over $\mathbb{Z}_p$, is called a splitting code \cite{agreferath1}. The lifting decoder technique is a process of lifting a decoding scheme for the linear code over $\mathbb{Z}_p$, defined by the generator matrix $G_0$ to decoding schemes for the corresponding linear codes over $\mathbb{Z}_{p^m}, m=\{1, \dots, k\}$ \cite{agreferath1, ababu1}. Instead of iteratively estimating the error element as done in \cite{agreferath1, ababu1}, we will define a residue that can easily be estimated at run-time and used to decode the codeword layer by layer.

\begin{figure}[bt]
\centering
\begin{tikzpicture}
[circuit ee IEC, node distance=14mm,auto,>=latex',
node distance=19mm,auto,>=latex',
  box/.style={draw, minimum size=0.6cm},
  short/.style={node distance=14mm},
  supershort/.style={node distance=10mm}]
\node (start) {};
\node[contact, supershort, right of = start] (I) {} edge[-] (start);
\node[contact, short, below of = I] (B) {} edge [-] (I);
\node[box, short, right of=I] (G0) {$G_0$} edge[<-] node[above,pos=1.5]
  {$\boldsymbol{u}_0$} (I);
\node[circle, short, draw=black!80,right of=G0] (A) {$+$} edge[<-] node[above,pos=0.3]
  {$\boldsymbol{r}_1$} (G0) ;
\node[box, short, below of = G0] (G1) {$G_1$} edge[<-] (B);
\node[contact, short, below of = A] (C) {} edge [->] (A) edge [<-] (G1);
\node[box, short, right of= A] (G01) {$G_0$} edge[->] (A);
\node [short, right of = G01] (end) {} edge [->] node[above,pos= 0.2]
  {$\boldsymbol{u}_1$} (G01);
  
\node[short, above of = G0] (c0) {$\boldsymbol{c}_0$} edge[<-] (G0);
\node[short, above of = A] (c1) {$\boldsymbol{c}_1$} edge[<-] (A);
\end{tikzpicture}
\caption{Graphical representation of the encoder, defined by eq. (\ref{eq:lifting_enco}).}
\label{fig:lifting_enco}
\end{figure}
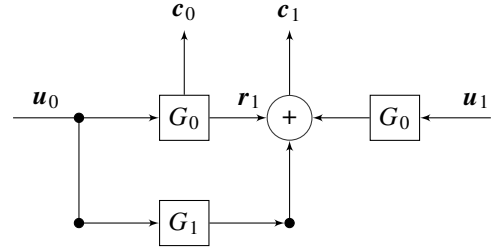

Consider again the codeword $\boldsymbol{c} = \boldsymbol{u} G \in \mathbb{Z}_4^N$, with a dyadic expansion $(\boldsymbol{c}_0, \boldsymbol{c}_1)$. By substituting $G = G_0 + 2\cdot G_1$ in eq. (\ref{eq:half_exp}), we obtain
\begin{equation}
\label{eq:cwdiad}
\boldsymbol{c} = \boldsymbol{u}_0 \cdot G_0 + 2 \cdot (\boldsymbol{u}_0 \cdot G_1 + \boldsymbol{u}_1 \cdot G_0).
\end{equation}
Let $(\boldsymbol{r}_0, \boldsymbol{r}_1)$ represent the dyadic expansion of the residue part $\boldsymbol{u}_0 \cdot G_0$, with
\begin{equation}
\boldsymbol{r}_0 = \boldsymbol{u}_0 \otimes G_0,
\end{equation}
and
\begin{equation}
\label{eq:r1}
\boldsymbol{r}_1 = \frac{\boldsymbol{u}_0 \cdot G_0 - \boldsymbol{u}_0 \otimes G_0}{2}.
\end{equation}
By substituting this back into (\ref{eq:cwdiad}), we get
\begin{equation}
\boldsymbol{c} = \boldsymbol{u}_0 \otimes G_0 + 2 \cdot (\boldsymbol{r}_1 + \boldsymbol{u}_0 \cdot G_1 + \boldsymbol{u}_1 \cdot G_0).
\end{equation}
Using \cite[][Lemma 1]{agreferath1}, we have
\begin{equation}
\label{eq:lifting_enco}
\boldsymbol{c} = \boldsymbol{u}_0 \otimes G_0 + 2 \cdot (\boldsymbol{r}_1 \oplus \boldsymbol{u}_0 \otimes G_1 \oplus \boldsymbol{u}_1 \otimes G_0) = \boldsymbol{c}_0 + 2 \cdot \boldsymbol{c}_1.
\end{equation}
A graphical representation of the encoder is given in Fig. \ref{fig:lifting_enco}.

Given the channel output $\boldsymbol{y}$, we begin the decoding process by calculating reliability elements $\boldsymbol{w}_0, \boldsymbol{w}_1 \in \mathbb{R}^N$ as 
\begin{equation}
\label{likely1}
w_{0, n} = \log \frac{P[y_n|c_{0, n} = 0]}{P[y_n|c_{0, n} = 1]}, \ n = 0, 1, \dots N-1,
\end{equation}
and
\begin{equation}
\label{likely2}
w_{1, n} = \log \frac{P[y_n|c_{1, n} = 0]}{P[y_n|c_{1, n} = 1]}, \ n = 0, 1, \dots N-1.
\end{equation}
Next, we compute $\boldsymbol{d}_0 \in \mathbb{R}^N$ as
\begin{equation}
\boldsymbol{d}_0 = D(\boldsymbol{w}_0),
\end{equation}
where $D(\cdot)$ represents the SISO decoding algorithm corresponding to the binary linear block code, defined by the generator matrix $G_0$. In the case of the Kerdock code, the corresponding binary code is the RM$(1,m)$ code, while in the case of the Preparata code, the corresponding binary code is the extended Hamming code. Low complexity SISO decoding algorithms for these codes were presented in \cite{aashikhmin1}.

Given $\boldsymbol{d}_0$ it is easy to find the hard decision estimates $\hat{\boldsymbol{c}}_0$ and $\hat{\boldsymbol{u}}_0$. We calculate $\hat{\boldsymbol{r}}_1$ using eq. (\ref{eq:r1}) and $\hat{\boldsymbol{p}}$ as
\begin{equation}
\label{eq:calcphat}
\hat{\boldsymbol{p}} = \hat{\boldsymbol{r}}_1 \oplus \hat{\boldsymbol{u}}_0 \otimes G_1.
\end{equation}
Let $\boldsymbol{b} = \phi(\hat{\boldsymbol{p}})$, where $\phi(c) = 1 - 2c$ represents the BPSK modulation. Finally, we compute
\begin{equation}
\boldsymbol{d}_1 = D(\boldsymbol{b} \odot \boldsymbol{w}_1).
\end{equation}

Elements $\boldsymbol{d}_0$ and $\boldsymbol{b} \cdot \boldsymbol{d}_1$ correspond to soft estimates of $\boldsymbol{c}_0$ and $\boldsymbol{c}_1$, respectively. This concludes the algorithm.

As the decoder $D(\cdot)$ of the corresponding binary subcode has complexity $\mathcal{O}(N \log_2 N)$ \cite{aashikhmin1} and assuming the calculation of $\hat{\boldsymbol{p}}$ (which has the same complexity as the encoding procedure) takes at most $\mathcal{O}(N \log_2 N)$, the lifting decoder also has complexity $\mathcal{O}(N \log_2 N)$. We summarize the APP decoding procedure in Algorithm \ref{alg2siso}. A detailed complexity analysis of Algorithm \ref{alg2siso} for the case of Kerdock and Preparata codes is given in Table \ref{minja3t}. Note that operations are now done using floating-point numbers with some binary and quaternary computations in step 4 and 5.

\begin{figure}[bt]
\begin{algorithm}[H]
\centering
\begin{algorithmic}[1]
 \renewcommand{\algorithmicrequire}{\textbf{Input:}}
 \renewcommand{\algorithmicensure}{\textbf{Output:}}
\REQUIRE $\boldsymbol{y}$,
\ENSURE $\boldsymbol{d}_0, \boldsymbol{d}_1$
\STATE Compute the reliability element $\boldsymbol{w}_0$ using eq. (\ref{likely1})
\STATE Compute the reliability element $\boldsymbol{w}_1$ using eq. (\ref{likely2})
\STATE $\boldsymbol{d}_0, \hat{\boldsymbol{u}}_0 \leftarrow D(\boldsymbol{w}_0)$
\STATE $\hat{\boldsymbol{r}}_1 \leftarrow (\boldsymbol{u}_0 \cdot G_0 - \boldsymbol{u}_0 \otimes G_0) / 2$ 
\COMMENT{Using eq. (\ref{eq:r1})}
\STATE $\hat{\boldsymbol{p}} \leftarrow \hat{\boldsymbol{r}}_1 \oplus \hat{\boldsymbol{u}}_0 \otimes G_1$ 
\COMMENT{Using eq. (\ref{eq:calcphat})}
\STATE $\boldsymbol{b} \leftarrow \phi(\hat{\boldsymbol{p}})$
\STATE $\boldsymbol{d}_1 \leftarrow D(\boldsymbol{b} \odot \boldsymbol{w}_1)$
\end{algorithmic}
\caption{Bitwise APP decoding algorithm}
\label{alg2siso}
\end{algorithm}
\end{figure}

\begin{table}[!tb]
\caption{{Computational cost of bitwise APP decoding}}
\label{minja3t}
\centering
\begin{tabular}{|c|c|c|} 
\hline
\parbox[t]{1.2cm}{\centering Steps} & {\centering Short description} & {\centering Complexity} \\ \hline \hline
1. - 2. & Compute $\boldsymbol{w}_0$ and $\boldsymbol{w}_1$ & $\mathcal{O}(N)$ \tabularnewline \hline 
3. & Decode $\boldsymbol{w}_0$ & $\mathcal{O}(N\log_2 N)$ \tabularnewline \hline
4. & Calculate $\boldsymbol{r}_1$ & $\mathcal{O}(N\log_2 N)$ \tabularnewline \hline 
5. & Calculate $\hat{\boldsymbol{p}}$ & $\mathcal{O}(N\log_2 N)$ \tabularnewline \hline 
6. & BPSK modulation of $\hat{\boldsymbol{p}}$  & $\mathcal{O}(N)$ \tabularnewline \hline 
7. & Decode $\boldsymbol{w}_1 \odot \boldsymbol{b}$ & $\mathcal{O}(N\log_2 N)$ \tabularnewline \hline 
\end{tabular}
\end{table}


\section{Simulation Results}
\label{sc_sim}

We will now demonstrate the error-correcting performance of the new decoders introduced in section \ref{sc_map} and \ref{sc_app}, for the case of four different Kerdock and Preparata codes of length\footnote{Note that the code length here is given as the number of quaternary symbols. The binary code length would be twice this.} $8, 32, 128$ and $512$, defined in terms of the irreducible monic polynomial ($h(Z)$) used to design an extension ring, as presented in Section \ref{sc_model}. These codes are the self-dual {Nordstrom-Robinson} code, defined by $h(Z) = 3 + Z + 2Z^2 + Z^3$, and Kerdock and Preparata codes over finite rings defined by $h(Z) = 3 + 2Z + 3Z^2 + Z^5$, $h(Z) = 3 + Z + 2Z^4 + Z^7$ and $h(Z) = 3 + 2Z^2 + 3Z^4 + Z^9$. All results are presented for the additive white Gaussian noise channel with a QPSK modulation. In the case of the MAP decoders, a standard QPSK modulation defined as $\phi(c) = I^c, c \in \mathbb{Z}_4$ is used, while in the case of the sub-optimal decoders we used a Gray-coded QPSK with respect to the dyadic expansion of a quaternary symbol. The performance of the proposed decoding algorithms is given in terms of the frame error rate (FER) and the symbol error rate (SER) as a function of the energy per bit to noise power spectral density ratio ($E_b/N_0$). In the case of the Nordstrom-Robinson code, we also provide the FER of a naive MAP decoder, implemented using eq. (\ref{eq:map_deco_rule}) (or eq. (\ref{eq:dualmap}), as the Nordstrom-Robinson is self-dual and both equations give the same result) and show that it matches the two novel MAP decoders. We also compared our novel decoding algorithms with the classical lifting decoder \cite{agreferath1} and the $\mathbb{Z}_4$ Chase decoder \cite{aarmond1,aarmond2} in terms of FER and SER. The original lifting decoder was implemented as a two stage decoder, where each stage used a binary hard decision decoder corresponding to the associated binary code. In the case of the Preparata codes we used the Meggitt decoder \cite{1057659} and in the case of the Kerdock codes we used the minimum Hamming distance decoder. The $\mathbb{Z}_4$ Chase decoder improves upon the original lifting decoder by using the Chase reprocessing step at every stage, i.e. at stage $i \in \{1, 2\}$, the decoder generates $2^{e_i}$ binary test patterns of maximum Hamming weight $e_i$, where ones are restricted to the $e_i$ least reliable positions in the corresponding channel output. Out of $2^{e_i}$ generated decoder outputs the one that minimizes the Euclidean distance is chosen as the correct one \cite{aarmond2}. In the case of the Preparata codes we used parameters $e_1 = 2$ and $e_2 = 1$, while in the case of the Kerdock codes we used parameters $e_1=8$ and $e_2=4$.

All simulations are done using the Monte-Carlo simulation with a relative precision $\delta = 0.05$ for a range of $E_b/N_0$ points, starting from $-1.6dB$ with a step of $0.5dB$. 
\begin{figure}[bt]
    \centering
%
%
\definecolor{mycolor1}{rgb}{0.00000,0.44700,0.74100}%
\definecolor{mycolor2}{rgb}{0.00000,0.49804,0.00000}%
\begin{tikzpicture}

\begin{axis}[%
width=\textwidth/2.6,
height=\textwidth/3.7,
at={(2.6in,1.04in)},
scale only axis,
xmin=-1.6,
xmax=8.4,
xlabel style={font=\color{white!15!black}},
xlabel={$\text{E}_\text{b}\text{/N}_\text{0}$},
ymode=log,
ymin=0.00023,
ymax=1,
yminorticks=true,
ylabel style={font=\color{white!15!black}},
ylabel={FER},
axis background/.style={fill=white},
legend style={at={(0.01,0.01)}, anchor=south west, legend cell align=left, align=left, draw=white!15!black}
]
\addplot [color=mycolor1, line width=1.0pt, mark=+, mark options={solid, mycolor1}]
  table[row sep=crcr]{%
-1.6	0.843464228079613\\
-1.1	0.804948453608247\\
-0.6	0.754531490015361\\
-0.1	0.713680538250685\\
0.4	0.650074294205052\\
0.9	0.578060805258833\\
1.4	0.502066740598851\\
1.9	0.430776214253291\\
2.4	0.359739311197258\\
2.9	0.280009332348252\\
3.4	0.208981640470215\\
3.9	0.152163379573964\\
4.4	0.107026858587917\\
4.9	0.0698364077269008\\
5.4	0.0410109476565173\\
5.9	0.0232310908481203\\
6.4	0.0117153789983784\\
6.9	0.00540735086563282\\
7.4	0.00225907961863258\\
7.9	0.000831077853580085\\
8.4	0.00026637838049831\\
8.9	7.41000079790889e-05\\
9.4	1.7352008958162e-05\\
};
\addlegendentry{\footnotesize Naive MAP decoding}

\addplot [color=red, dashed, line width=1.0pt, mark=o, mark options={solid, red}]
  table[row sep=crcr]{%
-1.6	0.855038302887448\\
-1.1	0.821527138914443\\
-0.6	0.763378465506125\\
-0.1	0.71561478501383\\
0.4	0.661712944857255\\
0.9	0.57758031442242\\
1.4	0.504219623792578\\
1.9	0.431476703596904\\
2.4	0.361271349394865\\
2.9	0.281703395812093\\
3.4	0.210520702802144\\
3.9	0.152093982602457\\
4.4	0.107052519270652\\
4.9	0.068231245194596\\
5.4	0.041509499330911\\
5.9	0.0231746054338982\\
6.4	0.0117694553872976\\
6.9	0.00545340399108462\\
7.4	0.00229794154827094\\
7.9	0.000834801462707175\\
8.4	0.000269332507143251\\
8.9	7.38232816217487e-05\\
};
\addlegendentry{\footnotesize $\mathcal{K}[8, 4]$ MAP}

\addplot [color=mycolor2, dotted, line width=1.0pt, mark=asterisk, mark options={solid, mycolor2}]
  table[row sep=crcr]{%
-1.6	0.851027397260274\\
-1.1	0.810051107325383\\
-0.6	0.770938446014127\\
-0.1	0.709779951100244\\
0.4	0.650716812511637\\
0.9	0.579702151130723\\
1.4	0.504374364191251\\
1.9	0.426184464577454\\
2.4	0.352989308313332\\
2.9	0.281576987890426\\
3.4	0.214266237661559\\
3.9	0.152039444195428\\
4.4	0.106358670427363\\
4.9	0.0690488904711048\\
5.4	0.0418493737006691\\
5.9	0.0229446449579783\\
6.4	0.0117403044187403\\
6.9	0.00546512343508229\\
7.4	0.00223400996953256\\
7.9	0.000813967924189543\\
8.4	0.000262653250737231\\
};
\addlegendentry{\footnotesize $\mathcal{P}[8, 4]$ MAP}

\end{axis}
\end{tikzpicture}%
    \caption{FER comparison of different MAP decoders for the Nordstrom-Robinson code.}
    \label{fig:fernr}
\end{figure}
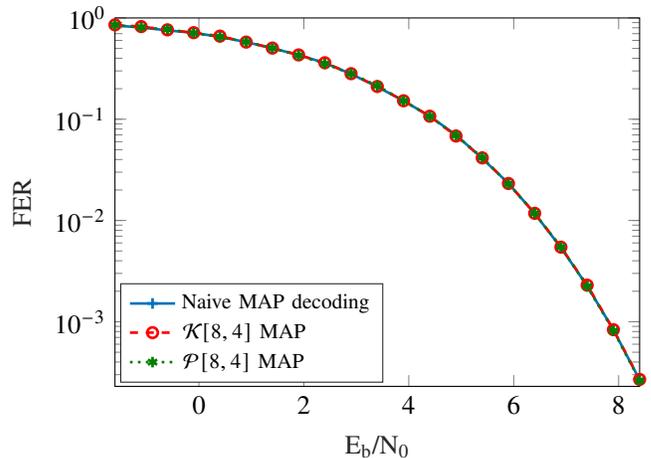

\begin{figure}[bt]
    \centering
%
%
\definecolor{mycolor1}{rgb}{0.00000,0.44700,0.74100}%
\definecolor{mycolor2}{rgb}{0.00000,0.49804,0.00000}%
\definecolor{mycolor3}{rgb}{0.85098,0.32549,0.09804}%
\definecolor{mycolor4}{rgb}{0.30196,0.74510,0.93333}%
\definecolor{mycolor5}{rgb}{0.74902,0.00000,0.74902}%
\begin{tikzpicture}

\begin{axis}[%
width=\textwidth/2.6,
height=\textwidth/3.7,
at={(1.293in,0.698in)},
scale only axis,
xmin=-1.6,
xmax=9.4,
xlabel style={font=\color{white!15!black}},
xlabel={$\text{E}_\text{b}\text{/N}_\text{0}$},
ymode=log,
ymin=0.0003,
ymax=1,
yminorticks=true,
ylabel style={font=\color{white!15!black}},
ylabel={FER},
axis background/.style={fill=white},
legend style={at={(0.01,0.01)}, anchor=south west, legend cell align=left, align=left, draw=white!15!black}
]
\addplot [color=mycolor1, line width=1.0pt, mark=+, mark options={solid, mycolor1}]
  table[row sep=crcr]{%
-1.6	0.851027397260274\\
-1.1	0.810051107325383\\
-0.6	0.770938446014127\\
-0.1	0.709779951100244\\
0.4	0.650716812511637\\
0.9	0.579702151130723\\
1.4	0.504374364191251\\
1.9	0.426184464577454\\
2.4	0.352989308313332\\
2.9	0.281576987890426\\
3.4	0.214266237661559\\
3.9	0.152039444195428\\
4.4	0.106358670427363\\
4.9	0.0690488904711048\\
5.4	0.0418493737006691\\
5.9	0.0229446449579783\\
6.4	0.0117403044187403\\
6.9	0.00546512343508229\\
7.4	0.00223400996953256\\
7.9	0.000813967924189543\\
8.4	0.000262653250737231\\
};
\addlegendentry{\footnotesize $\mathcal{P}[8, 4]$ MAP}

\addplot [color=red, dashed, line width=1.0pt, mark=o, mark options={solid, red}]
  table[row sep=crcr]{%
-1.6	0.861214374225527\\
-1.1	0.81389252948886\\
-0.6	0.767756854971919\\
-0.1	0.713220675944334\\
0.4	0.643294881038212\\
0.9	0.586177595241467\\
1.4	0.507309038501132\\
1.9	0.433992640294388\\
2.4	0.347347134266404\\
2.9	0.281599059376837\\
3.4	0.212831535404331\\
3.9	0.155613702113545\\
4.4	0.107504606824122\\
4.9	0.0688512953980835\\
5.4	0.0409135700299888\\
5.9	0.0230504443757624\\
6.4	0.0117880699912928\\
6.9	0.00543630353223084\\
7.4	0.0022194694493745\\
7.9	0.000834033299731051\\
8.4	0.000261146690468506\\
};
\addlegendentry{\footnotesize $\mathcal{P}[8, 4]$ ML}

\addplot [color=mycolor2, line width=1.0pt, mark=+, mark options={solid, mycolor2}]
  table[row sep=crcr]{%
-1.6	1\\
-1.1	1\\
-0.6	1\\
-0.1	1\\
0.4	1\\
0.9	1\\
1.4	1\\
1.9	0.99009900990099\\
2.4	0.99009900990099\\
2.9	0.95136186770428\\
3.4	0.897637795275591\\
3.9	0.771515356058049\\
4.4	0.616522018643523\\
4.9	0.433667610809156\\
5.4	0.255198876289013\\
5.9	0.128719160880485\\
6.4	0.0503704725182054\\
6.9	0.0161950850163778\\
7.4	0.00419338145539264\\
7.9	0.000843546376351187\\
8.4	0.000134575816662489\\
};
\addlegendentry{\footnotesize $\mathcal{P}[32, 26]$ MAP}

\addplot [color=mycolor3, dashed, line width=1.0pt, mark=o, mark options={solid, mycolor3}]
  table[row sep=crcr]{%
-1.6	1\\
-1.1	1\\
-0.6	1\\
-0.1	1\\
0.4	1\\
0.9	1\\
1.4	1\\
1.9	1\\
2.4	1\\
2.9	0.980392156862745\\
3.4	0.91743119266055\\
3.9	0.813008130081301\\
4.4	0.637554585152838\\
4.9	0.431261770244821\\
5.4	0.228550295857988\\
5.9	0.122010398613518\\
6.4	0.047852298417483\\
6.9	0.0166836043360434\\
7.4	0.00405817737998373\\
7.9	0.000828052771803147\\
8.4	0.000164536741214058\\
};
\addlegendentry{\footnotesize $\mathcal{P}[32, 26]$ ML}

\addplot [color=mycolor4, line width=1.0pt, mark=+, mark options={solid, mycolor4}]
  table[row sep=crcr]{%
-1.6	1\\
-1.1	1\\
-0.6	1\\
-0.1	1\\
0.4	1\\
0.9	1\\
1.4	1\\
1.9	1\\
2.4	1\\
2.9	1\\
3.4	1\\
3.9	1\\
4.4	1\\
4.9	0.99009900990099\\
5.4	0.99009900990099\\
5.9	0.933147632311978\\
6.4	0.707317073170732\\
6.9	0.370859365790404\\
7.4	0.127081635029696\\
7.9	0.0270707902556284\\
8.4	0.00345057414354541\\
8.9	0.000290423713677069\\
};
\addlegendentry{\footnotesize $\mathcal{P}[128, 120]$ MAP}

\addplot [color=mycolor5, dashed, line width=1.0pt, mark=o, mark options={solid, mycolor5}]
  table[row sep=crcr]{%
-1.6	1\\
-1.1	1\\
-0.6	1\\
-0.1	1\\
0.4	1\\
0.9	1\\
1.4	1\\
1.9	1\\
2.4	1\\
2.9	1\\
3.4	1\\
3.9	1\\
4.4	1\\
4.9	1\\
5.4	0.99009900990099\\
5.9	0.943396226415094\\
6.4	0.657142857142857\\
6.9	0.378995433789954\\
7.4	0.126811594202899\\
7.9	0.0253049571762263\\
8.4	0.00363523720150512\\
8.9	0.000277408285353259\\
};
\addlegendentry{\footnotesize $\mathcal{P}[128, 120]$ ML}

\addplot [color=green, line width=1.0pt, mark=+, mark options={solid, green}]
  table[row sep=crcr]{%
-1.6	1\\
-1.1	1\\
-0.6	1\\
-0.1	1\\
0.4	1\\
0.9	1\\
1.4	1\\
1.9	1\\
2.4	1\\
2.9	1\\
3.4	1\\
3.9	1\\
4.4	1\\
4.9	1\\
5.4	1\\
5.9	1\\
6.4	1\\
6.9	1\\
7.4	0.99009900990099\\
7.9	0.804437140509449\\
8.4	0.334758981583979\\
8.9	0.05364562372107\\
9.4	0.00331059889480759\\
};
\addlegendentry{\footnotesize $\mathcal{P}[512, 502]$ MAP}

\addplot [color=blue, dashed, line width=1.0pt, mark=o, mark options={solid, blue}]
  table[row sep=crcr]{%
-1.6	1\\
-1.1	1\\
-0.6	1\\
-0.1	1\\
0.4	1\\
0.9	1\\
1.4	1\\
1.9	1\\
2.4	1\\
2.9	1\\
3.4	1\\
3.9	1\\
4.4	1\\
4.9	1\\
5.4	1\\
5.9	1\\
6.4	1\\
6.9	1\\
7.4	1\\
7.9	0.78740157480315\\
8.4	0.349865951742627\\
8.9	0.049429164504411\\
9.4	0.0035\\
};
\addlegendentry{\footnotesize $\mathcal{P}[512, 502]$ ML}

\end{axis}
\end{tikzpicture}%
    \caption{FER comparison of the novel MAP decoding algorithm with an ML lower bound for different Preparata codes.}
    \label{fig:ferml}
\end{figure}
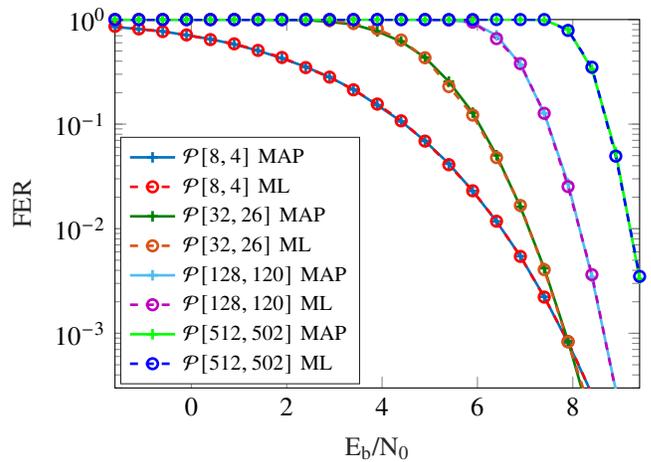

\begin{figure}[bt]
    \centering
%
%
\definecolor{mycolor1}{rgb}{0.00000,0.44700,0.74100}%
\definecolor{mycolor2}{rgb}{0.85000,0.32500,0.09800}%
\definecolor{mycolor3}{rgb}{0.00000,0.49804,0.00000}%
\definecolor{mycolor4}{rgb}{0.00000,0.44706,0.74118}%
\begin{tikzpicture}

\begin{axis}[%
width=\textwidth/2.6,
height=\textwidth/3.5,
at={(1.293in,0.731in)},
scale only axis,
xmin=-1.6,
xmax=9.4,
xlabel style={font=\color{white!15!black}},
xlabel={$\text{E}_\text{b}\text{/N}_\text{0}$},
ymode=log,
ymin=0.0006,
ymax=1,
yminorticks=true,
ylabel style={font=\color{white!15!black}},
ylabel={FER},
axis background/.style={fill=white},
legend style={at={(0.01,0.01)}, anchor=south west, legend cell align=left, align=left, draw=white!15!black}
]
\addplot [color=mycolor1, dashed, line width=1.0pt, mark=+, mark options={solid, mycolor1}]
  table[row sep=crcr]{%
-1.6	0.901459854014599\\
-1.1	0.872689938398357\\
-0.6	0.804018040180402\\
-0.1	0.732675979183785\\
0.4	0.670117742590337\\
0.9	0.56695902605053\\
1.4	0.47270771712826\\
1.9	0.375910870219813\\
2.4	0.280649061902719\\
2.9	0.195207721970267\\
3.4	0.132738486967967\\
3.9	0.0801662759588319\\
4.4	0.0428482539739383\\
4.9	0.0210659499796657\\
5.4	0.00906149104046216\\
5.9	0.00338965527678557\\
6.4	0.00109098569520088\\
6.9	0.000287423037149039\\
7.4	6.39248807986355e-05\\
};
\addlegendentry{\footnotesize $\mathcal{K} [32, 6]$}

\addplot [color=mycolor2, dashed, line width=1.0pt, mark=o, mark options={solid, mycolor2}]
  table[row sep=crcr]{%
-1.6	0.923821039903265\\
-1.1	0.865892972275951\\
-0.6	0.816444444444444\\
-0.1	0.749476831091181\\
0.4	0.666134822497008\\
0.9	0.554973821989529\\
1.4	0.437077866625262\\
1.9	0.331761318506751\\
2.4	0.230880057629968\\
2.9	0.146160692889665\\
3.4	0.0837086504115546\\
3.9	0.0427837839045733\\
4.4	0.0184092571407138\\
4.9	0.00698972901867189\\
5.4	0.0021756197594849\\
5.9	0.000528563642711496\\
};
\addlegendentry{\footnotesize $\mathcal{K} [128, 8]$}

\addplot [color=mycolor3, dashed, line width=1.0pt, mark=asterisk, mark options={solid, mycolor3}]
  table[row sep=crcr]{%
-1.6	0.938066465256798\\
-1.1	0.900542495479204\\
-0.6	0.855621301775148\\
-0.1	0.777506112469438\\
0.4	0.671431488666531\\
0.9	0.556195965417867\\
1.4	0.429043648110585\\
1.9	0.304513417677188\\
2.4	0.196896526364817\\
2.9	0.116206302670155\\
3.4	0.05945826353551\\
3.9	0.0253077460345486\\
4.4	0.00932253140777269\\
4.9	0.00269875410900532\\
5.4	0.000581997648729499\\
};
\addlegendentry{\footnotesize $\mathcal{K} [512, 10]$}

\addplot [color=mycolor4, line width=1.0pt, mark=+, mark options={solid, mycolor4}]
  table[row sep=crcr]{%
-1.6	1\\
-1.1	1\\
-0.6	1\\
-0.1	1\\
0.4	1\\
0.9	1\\
1.4	1\\
1.9	0.99009900990099\\
2.4	0.99009900990099\\
2.9	0.95136186770428\\
3.4	0.897637795275591\\
3.9	0.771515356058049\\
4.4	0.616522018643523\\
4.9	0.433667610809156\\
5.4	0.255198876289013\\
5.9	0.128719160880485\\
6.4	0.0503704725182054\\
6.9	0.0161950850163778\\
7.4	0.00419338145539264\\
7.9	0.000843546376351187\\
8.4	0.000134575816662489\\
};
\addlegendentry{\footnotesize $\mathcal{P} [32, 26]$}

\addplot [color=red, line width=1.0pt, mark=o, mark options={solid, red}]
  table[row sep=crcr]{%
-1.6	1\\
-1.1	1\\
-0.6	1\\
-0.1	1\\
0.4	1\\
0.9	1\\
1.4	1\\
1.9	1\\
2.4	1\\
2.9	1\\
3.4	1\\
3.9	1\\
4.4	1\\
4.9	0.99009900990099\\
5.4	0.99009900990099\\
5.9	0.933147632311978\\
6.4	0.707317073170732\\
6.9	0.370859365790404\\
7.4	0.127081635029696\\
7.9	0.0270707902556284\\
8.4	0.00345057414354541\\
8.9	0.000290423713677069\\
};
\addlegendentry{\footnotesize $\mathcal{P} [128, 120]$}

\addplot [color=mycolor3, line width=1.0pt, mark=asterisk, mark options={solid, mycolor3}]
  table[row sep=crcr]{%
-1.6	1\\
-1.1	1\\
-0.6	1\\
-0.1	1\\
0.4	1\\
0.9	1\\
1.4	1\\
1.9	1\\
2.4	1\\
2.9	1\\
3.4	1\\
3.9	1\\
4.4	1\\
4.9	1\\
5.4	1\\
5.9	1\\
6.4	1\\
6.9	1\\
7.4	0.99009900990099\\
7.9	0.804437140509449\\
8.4	0.334758981583979\\
8.9	0.05364562372107\\
9.4	0.00331059889480759\\
};
\addlegendentry{\footnotesize $\mathcal{P} [512, 502]$}

\end{axis}
\end{tikzpicture}%
    \caption{FER comparison of different Kerdock and Preparata codes using the novel MAP decoding algorithms.}
    \label{fig:fermap}
\end{figure}
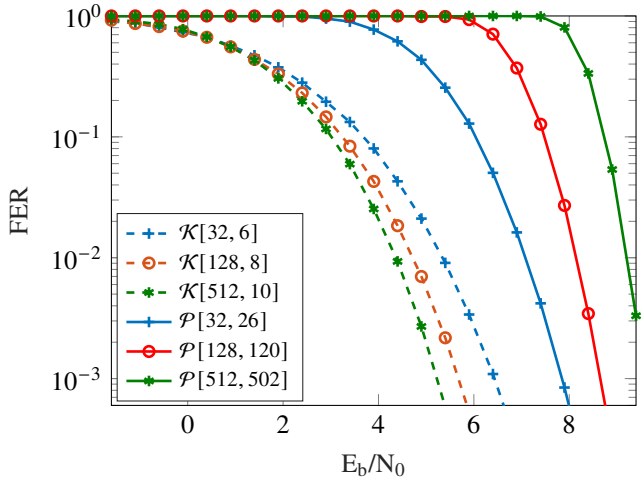
FER of the Nordstrom-Robinson code using the naive MAP decoder compared to the two novel MAP decoders is presented in Fig. \ref{fig:fernr}. As expected, all three error-rate curves overlap. If the Euclidean distance between the correct codeword and the channel output is greater than the distance between the decoded sequence and the channel output, we assume that the Maximum-Likelihood (ML) decoder would also make a mistake. This allows us to simulate the ML lower bound. In Fig. \ref{fig:ferml} it is shown that the FER performance of the MAP decoder perfectly coincides with the ML lower bound, for several different Preparata codes, as expected. FER of different Kerdock and Preparata codes using novel MAP decoders is presented in Fig. \ref{fig:fermap}. SER of different Kerdock and Preparata codes using novel MAP decoders is presented in Fig. \ref{fig:sermap}. 
\begin{figure}[bt]
    \centering
%
%
\definecolor{mycolor1}{rgb}{0.00000,0.44700,0.74100}%
\definecolor{mycolor2}{rgb}{0.85000,0.32500,0.09800}%
\definecolor{mycolor3}{rgb}{0.00000,0.49804,0.00000}%
\definecolor{mycolor4}{rgb}{0.00000,0.44706,0.74118}%
\begin{tikzpicture}

\begin{axis}[%
width=\textwidth/2.6,
height=\textwidth/3.5,
at={(1.293in,0.731in)},
scale only axis,
xmin=-1.6,
xmax=8.4,
xlabel style={font=\color{white!15!black}},
xlabel={$\text{E}_\text{b}\text{/N}_\text{0}$},
ymode=log,
ymin=0.0007,
ymax=1,
yminorticks=true,
ylabel style={font=\color{white!15!black}},
ylabel={SER},
axis background/.style={fill=white},
legend style={at={(0.01,0.01)}, anchor=south west, legend cell align=left, align=left, draw=white!15!black}
]
\addplot [color=mycolor1, dashed, line width=1.0pt, mark=+, mark options={solid, mycolor1}]
  table[row sep=crcr]{%
-1.6	0.531867687152107\\
-1.1	0.497139715330632\\
-0.6	0.459112174318934\\
-0.1	0.4195834839176\\
0.4	0.371145845627287\\
0.9	0.321284671273726\\
1.4	0.26215271609763\\
1.9	0.210280824835592\\
2.4	0.158182390409801\\
2.9	0.112692881635763\\
3.4	0.0751192893401015\\
3.9	0.0457362590224591\\
4.4	0.0254009323325906\\
4.9	0.0125722209419724\\
5.4	0.00560380108308538\\
5.9	0.00211306762261063\\
6.4	0.000682351094393958\\
6.9	0.000185955167170581\\
};
\addlegendentry{\footnotesize $\mathcal{K}[32, 6]$}

\addplot [color=mycolor2, dashed, line width=1.0pt, mark=o, mark options={solid, mycolor2}]
  table[row sep=crcr]{%
-1.6	0.592378868226064\\
-1.1	0.552130839393566\\
-0.6	0.513497091336007\\
-0.1	0.460343150319829\\
0.4	0.397050805681743\\
0.9	0.332789900249377\\
1.4	0.262001204696319\\
1.9	0.195809340913518\\
2.4	0.134736276879837\\
2.9	0.0878653594236122\\
3.4	0.0498143020969856\\
3.9	0.0260491051727873\\
4.4	0.0114921232866759\\
4.9	0.00432811889395593\\
5.4	0.00133231203565379\\
5.9	0.000346289651195786\\
6.4	6.83852788685616e-05\\
};
\addlegendentry{\footnotesize $\mathcal{K}[128, 8]$}

\addplot [color=mycolor3, dashed, line width=1.0pt, mark=asterisk, mark options={solid, mycolor3}]
  table[row sep=crcr]{%
-1.6	0.629034383476013\\
-1.1	0.598878847947761\\
-0.6	0.547324086526838\\
-0.1	0.491543442543732\\
0.4	0.426938226311085\\
0.9	0.344166821082804\\
1.4	0.269222180271107\\
1.9	0.187505002696949\\
2.4	0.122791942458568\\
2.9	0.0685374540441176\\
3.4	0.0347075259340758\\
3.9	0.0158537811745776\\
4.4	0.00588099780134506\\
4.9	0.00169075234678726\\
5.4	0.000402641053277551\\
};
\addlegendentry{\footnotesize $\mathcal{K}[512, 10]$}

\addplot [color=mycolor4, line width=1.0pt, mark=+, mark options={solid, mycolor4}]
  table[row sep=crcr]{%
-1.6	0.401495841159109\\
-1.1	0.381975774055992\\
-0.6	0.360937764725815\\
-0.1	0.338743085079137\\
0.4	0.314914563798492\\
0.9	0.291942383013812\\
1.4	0.268276615830187\\
1.9	0.244029414138154\\
2.4	0.219600298263879\\
2.9	0.194026684480393\\
3.4	0.165875183940503\\
3.9	0.134139575684343\\
4.4	0.100243081473724\\
4.9	0.0679685392070356\\
5.4	0.0397574950522925\\
5.9	0.0197851741475289\\
6.4	0.0081153410955015\\
6.9	0.00268423132811325\\
7.4	0.000707021488203467\\
};
\addlegendentry{\footnotesize $\mathcal{P}[32, 26]$}

\addplot [color=red, line width=1.0pt, mark=o, mark options={solid, red}]
  table[row sep=crcr]{%
-1.6	0.376452139299728\\
-1.1	0.354851544024638\\
-0.6	0.33249839679733\\
-0.1	0.309610776345291\\
0.4	0.286703144467666\\
0.9	0.262823919340463\\
1.4	0.239221825477927\\
1.9	0.215226404810224\\
2.4	0.191428259647254\\
2.9	0.168606272306944\\
3.4	0.146257724971734\\
3.9	0.12508790388877\\
4.4	0.105110332980973\\
4.9	0.086719002217137\\
5.4	0.0696695905036844\\
5.9	0.0528813740899142\\
6.4	0.0345277670738816\\
6.9	0.017155392434885\\
7.4	0.00579019228529416\\
7.9	0.00121292033611711\\
8.4	0.000165946787081945\\
};
\addlegendentry{\footnotesize $\mathcal{P}[128, 120]$}

\addplot [color=mycolor3, line width=1.0pt, mark=asterisk, mark options={solid, mycolor3}]
  table[row sep=crcr]{%
-1.6	0.368179653813658\\
-1.1	0.346349682433506\\
-0.6	0.32410955106689\\
-0.1	0.300701616148069\\
0.4	0.278106219951923\\
0.9	0.253071793041455\\
1.4	0.230018365205224\\
1.9	0.205976840975373\\
2.4	0.182859008174958\\
2.9	0.159830419736217\\
3.4	0.138373263192215\\
3.9	0.117451207598936\\
4.4	0.0977710152653669\\
4.9	0.080128292337832\\
5.4	0.0642266174515354\\
5.9	0.0500887620271517\\
6.4	0.0381814813206986\\
6.9	0.0282077902789871\\
7.4	0.0200956507599836\\
7.9	0.0121918750964209\\
8.4	0.00441091797870102\\
};
\addlegendentry{\footnotesize $\mathcal{P}[512, 502]$}

\end{axis}
\end{tikzpicture}%
    \caption{SER comparison of different Kerdock and Preparata codes using the novel MAP decoding algorithms.}
    \label{fig:sermap}
\end{figure}
\begin{figure}[bt]
    \centering
%
%
\definecolor{mycolor1}{rgb}{0.00000,0.44700,0.74100}%
\definecolor{mycolor2}{rgb}{0.00000,0.44706,0.74118}%
\definecolor{mycolor3}{rgb}{0.00000,0.49804,0.00000}%
\begin{tikzpicture}

\begin{axis}[%
width=\textwidth/2.6,
height=\textwidth/3.5,
at={(2.6in,1.04in)},
scale only axis,
xmin=-1.6,
xmax=10,
xlabel style={font=\color{white!15!black}},
xlabel={$\text{E}_\text{b}\text{/N}_\text{0}$},
ymode=log,
ymin=0.0005,
ymax=1,
yminorticks=true,
ylabel style={font=\color{white!15!black}},
ylabel={FER},
axis background/.style={fill=white},
legend style={at={(0.01,0.01)}, anchor=south west, legend cell align=left, align=left, draw=white!15!black}
]
\addplot [color=mycolor1, dashed, line width=1.0pt, mark=+, mark options={solid, mycolor1}]
  table[row sep=crcr]{%
-1.6	0.843464228079613\\
-1.1	0.804948453608247\\
-0.6	0.754531490015361\\
-0.1	0.713680538250685\\
0.4	0.650074294205052\\
0.9	0.578060805258833\\
1.4	0.502066740598851\\
1.9	0.430776214253291\\
2.4	0.359739311197258\\
2.9	0.280009332348252\\
3.4	0.208981640470215\\
3.9	0.152163379573964\\
4.4	0.107026858587917\\
4.9	0.0698364077269008\\
5.4	0.0410109476565173\\
5.9	0.0232310908481203\\
6.4	0.0117153789983784\\
6.9	0.00540735086563282\\
7.4	0.00225907961863258\\
7.9	0.000831077853580085\\
8.4	0.00026637838049831\\
8.9	7.41000079790889e-05\\
9.4	1.7352008958162e-05\\
};
\addlegendentry{\footnotesize NR MAP}

\addplot [color=mycolor2, line width=1.0pt, mark=+, mark options={solid, mycolor2}]
  table[row sep=crcr]{%
-1.6	0.8810051736881\\
-1.1	0.827272727272727\\
-0.6	0.800079968012795\\
-0.1	0.759886111989877\\
0.4	0.695324971493729\\
0.9	0.657702650787553\\
1.4	0.573735199138859\\
1.9	0.508213658435789\\
2.4	0.424059486122359\\
2.9	0.369435332708529\\
3.4	0.295221342714746\\
3.9	0.23061707675248\\
4.4	0.176876302614899\\
4.9	0.12842009105509\\
5.4	0.0866718539464445\\
5.9	0.0567927993016045\\
6.4	0.0354383502760929\\
6.9	0.0208320479954047\\
7.4	0.0111685343006932\\
7.9	0.00577115063150074\\
8.4	0.00258640136281961\\
8.9	0.00107701532854789\\
9.4	0.000415339972761908\\
};
\addlegendentry{\footnotesize NR Lifting}

\addplot [color=red, dashed, line width=1.0pt, mark=o, mark options={solid, red}]
  table[row sep=crcr]{%
-1.6	0.923821039903265\\
-1.1	0.865892972275951\\
-0.6	0.816444444444444\\
-0.1	0.749476831091181\\
0.4	0.666134822497008\\
0.9	0.554973821989529\\
1.4	0.437077866625262\\
1.9	0.331761318506751\\
2.4	0.230880057629968\\
2.9	0.146160692889665\\
3.4	0.0837086504115546\\
3.9	0.0427837839045733\\
4.4	0.0184092571407138\\
4.9	0.00698972901867189\\
5.4	0.0021756197594849\\
5.9	0.000528563642711496\\
};
\addlegendentry{\footnotesize $\mathcal{K}[128, 8]$ MAP}

\addplot [color=red, line width=1.0pt, mark=o, mark options={solid, red}]
  table[row sep=crcr]{%
-1.6	0.980392156862745\\
-1.1	0.934579439252336\\
-0.6	0.952380952380952\\
-0.1	0.806451612903226\\
0.4	0.775193798449612\\
0.9	0.688524590163934\\
1.4	0.627615062761506\\
1.9	0.552147239263804\\
2.4	0.426716141001855\\
2.9	0.316397228637413\\
3.4	0.246732026143791\\
3.9	0.153324287652646\\
4.4	0.0978907517577069\\
4.9	0.0576307363927428\\
5.4	0.0291015186653699\\
5.9	0.0124602750070797\\
6.4	0.00515430539907015\\
6.9	0.00179505014921354\\
7.4	0.000610213757879385\\
7.9	0.000134251524258244\\
};
\addlegendentry{\footnotesize $\mathcal{K}[128, 8]$ Lifting}

\addplot [color=mycolor3, dashed, line width=1.0pt, mark=asterisk, mark options={solid, mycolor3}]
  table[row sep=crcr]{%
-1.6	1\\
-1.1	1\\
-0.6	1\\
-0.1	1\\
0.4	1\\
0.9	1\\
1.4	1\\
1.9	1\\
2.4	1\\
2.9	1\\
3.4	1\\
3.9	1\\
4.4	1\\
4.9	0.99009900990099\\
5.4	0.99009900990099\\
5.9	0.933147632311978\\
6.4	0.707317073170732\\
6.9	0.370859365790404\\
7.4	0.127081635029696\\
7.9	0.0270707902556284\\
8.4	0.00345057414354541\\
8.9	0.000290423713677069\\
};
\addlegendentry{\footnotesize $\mathcal{P}[128, 120]$ MAP}

\addplot [color=mycolor3, line width=1.0pt, mark=asterisk, mark options={solid, mycolor3}]
  table[row sep=crcr]{%
-1.6	1\\
-1.1	1\\
-0.6	1\\
-0.1	1\\
0.4	1\\
0.9	1\\
1.4	1\\
1.9	1\\
2.4	1\\
2.9	1\\
3.4	1\\
3.9	1\\
4.4	1\\
4.9	1\\
5.4	0.99009900990099\\
5.9	0.934579439252336\\
6.4	0.793650793650794\\
6.9	0.521739130434783\\
7.4	0.269797421731123\\
7.9	0.0897908979089791\\
8.4	0.026537833424061\\
8.9	0.00535659008627071\\
9.4	0.00100252385380195\\
9.9	0.00012412692226055\\
};
\addlegendentry{\footnotesize $\mathcal{P}[128, 120]$ Lifting}

\end{axis}
\end{tikzpicture}%
    \caption{FER of different Kerdock and Preparata codes using a sub-optimal decoder compared to the optimal MAP decoding.}
    \label{fig:ferlift}
\end{figure}
Sub-optimal decoding of different Kerdock and Preparata codes compared to optimal decoding is presented in Fig. \ref{fig:ferlift}. Note that Kerdock and Preparata families are not fixed-rate sequences of codes, i.e. the code rate of Kerdock codes decreases with length, while the code rate of Preparata codes increases. It follows that the error-correcting performance of Kerdock codes increases with length, while the performance of Preparata codes decreases with length, as seen in Fig. \ref{fig:fermap} and Fig. \ref{fig:sermap}. 

\begin{figure}[bt]
    \centering
%
%
\definecolor{mycolor1}{rgb}{0.00000,0.44700,0.74100}%
\definecolor{mycolor2}{rgb}{0.85000,0.32500,0.09800}%
\definecolor{mycolor3}{rgb}{0.00000,0.49804,0.00000}%
\definecolor{mycolor4}{rgb}{00.85000,0.32500,0.09800}%
\begin{tikzpicture}

\begin{axis}[%
width=\textwidth/2.6,
height=\textwidth/3.5,
at={(1.293in,0.731in)},
scale only axis,
xmin=-1.6,
xmax=10.5,
xlabel style={font=\color{white!15!black}},
xlabel={$\text{E}_\text{b}\text{/N}_\text{0}$},
ymode=log,
ymin=0.00034,
ymax=1,
yminorticks=true,
ylabel style={font=\color{white!15!black}},
ylabel={SER},
axis background/.style={fill=white},
legend style={at={(0.01,0.01)}, anchor=south west, legend cell align=left, align=left, draw=white!15!black}
]
\addplot [color=mycolor1, line width=1.0pt, mark=+, mark options={solid, mycolor1}]
  table[row sep=crcr]{%
-1.6	0.376452139299728\\
-1.1	0.354851544024638\\
-0.6	0.33249839679733\\
-0.1	0.309610776345291\\
0.4	0.286703144467666\\
0.9	0.262823919340463\\
1.4	0.239221825477927\\
1.9	0.215226404810224\\
2.4	0.191428259647254\\
2.9	0.168606272306944\\
3.4	0.146257724971734\\
3.9	0.12508790388877\\
4.4	0.105110332980973\\
4.9	0.086719002217137\\
5.4	0.0696695905036844\\
5.9	0.0528813740899142\\
6.4	0.0345277670738816\\
6.9	0.017155392434885\\
7.4	0.00579019228529416\\
7.9	0.00121292033611711\\
8.4	0.000165946787081945\\
};
\addlegendentry{\footnotesize  MAP}

\addplot [color=red, line width=1.0pt, mark=x, mark options={solid, mycolor2}]
  table[row sep=crcr]{%
-1.6	0.374226485148515\\
-1.1	0.353883044554455\\
-0.6	0.329285272277228\\
-0.1	0.305693069306931\\
0.4	0.2890625\\
0.9	0.267326732673267\\
1.4	0.235457920792079\\
1.9	0.214031559405941\\
2.4	0.186417079207921\\
2.9	0.172261757425743\\
3.4	0.148824257425743\\
3.9	0.122292698019802\\
4.4	0.109993811881188\\
4.9	0.0841584158415842\\
5.4	0.0735612623762376\\
5.9	0.0517669392523364\\
6.4	0.0385044642857143\\
6.9	0.0217179008152174\\
7.4	0.0104814111418048\\
7.9	0.00343250307503075\\
8.4	0.00100686411268372\\
8.9	0.000205771961346415\\
9.4	3.77708694137491e-05\\
9.9	4.70809537167945e-06\\
};
\addlegendentry{\footnotesize  SISO lifting}

\addplot [color=mycolor3, dashed, line width=1.0pt, mark=o, mark options={solid, mycolor3}]
  table[row sep=crcr]{%
-1.6	0.3832310892504\\
-1.1	0.371857219788978\\
-0.6	0.331101309649078\\
-0.1	0.315437644157376\\
0.4	0.30892623106608\\
0.9	0.26626320209452\\
1.4	0.259176693792733\\
1.9	0.205901445470943\\
2.4	0.205634223107729\\
2.9	0.17810225682221\\
3.4	0.152779542645442\\
3.9	0.139659415815423\\
4.4	0.116642700365169\\
4.9	0.111083149404896\\
5.4	0.0911081255451167\\
5.9	0.0720389880290229\\
6.4	0.0496491165507128\\
6.9	0.0342287892251499\\
7.4	0.0207458148923147\\
7.9	0.00945089688082787\\
8.4	0.00360734536265775\\
8.9	0.00116212830238992\\
9.4	0.000327500080563727\\
};
\addlegendentry{\footnotesize  Chase}

\addplot [color=mycolor4, dashed, line width=1.0pt, mark=asterisk, mark options={solid, mycolor4}]
  table[row sep=crcr]{%
-1.6	0.381574876237624\\
-1.1	0.373607673267327\\
-0.6	0.341584158415842\\
-0.1	0.325030940594059\\
0.4	0.308245668316832\\
0.9	0.273360148514851\\
1.4	0.248762376237624\\
1.9	0.223855198019802\\
2.4	0.211401608910891\\
2.9	0.186881188118812\\
3.4	0.162902227722772\\
3.9	0.143177599009901\\
4.4	0.125309405940594\\
4.9	0.111154084158416\\
5.4	0.0851639851485149\\
5.9	0.0712890625\\
6.4	0.0587418542654028\\
6.9	0.0454654211051931\\
7.4	0.032762368562955\\
7.9	0.0219518111254851\\
8.4	0.0130185208159214\\
8.9	0.00685131195335277\\
9.4	0.00294043864678899\\
9.9	0.00106751530776668\\
10.4	0.000334854590699747\\
10.9	0.000101792857994249\\
};
\addlegendentry{\footnotesize  Classical lifiting}

\end{axis}
\end{tikzpicture}%
    \caption{SER  comparison of the Preparata $\mathcal{P}[128,120]$ code decoded using MAP, SISO lifting, Chase and classical lifting decoders.}
    \label{fig:serp128}
\end{figure}
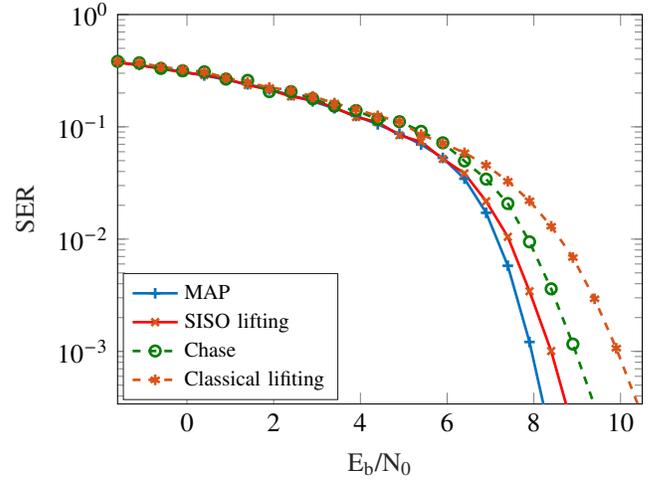

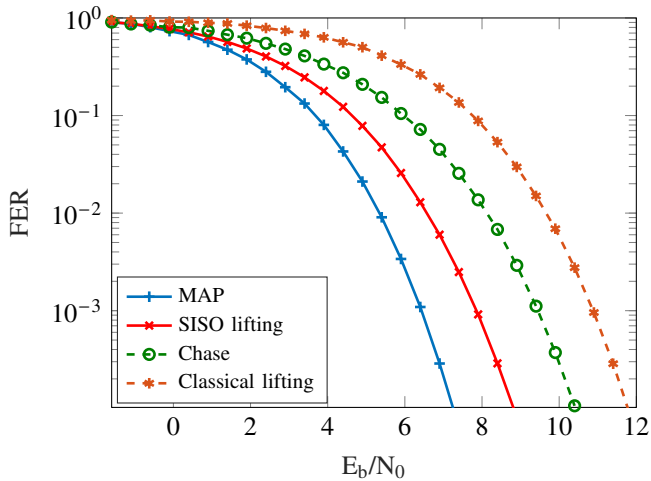
\begin{figure}[bt]
    \centering
%
%
\definecolor{mycolor1}{rgb}{0.00000,0.44706,0.74118}%
\definecolor{mycolor2}{rgb}{0.00000,0.49804,0.00000}%
\definecolor{mycolor3}{rgb}{0.85098,0.32549,0.09804}%
\begin{tikzpicture}

\begin{axis}[%
width=\textwidth/2.6,
height=\textwidth/3.5,
at={(1.293in,0.731in)},
scale only axis,
xmin=-1.6,
xmax=12,
xlabel style={font=\color{white!15!black}},
xlabel={$\text{E}_\text{b}\text{/N}_\text{0}$},
ymode=log,
ymin=0.000102,
ymax=1,
yminorticks=true,
ylabel style={font=\color{white!15!black}},
ylabel={FER},
axis background/.style={fill=white},
legend style={at={(0.01,0.01)}, anchor=south west, legend cell align=left, align=left, draw=white!15!black}
]
\addplot [color=mycolor1, line width=1.0pt, mark=+, mark options={solid, mycolor1}]
  table[row sep=crcr]{%
-1.6	0.901459854014599\\
-1.1	0.872689938398357\\
-0.6	0.804018040180402\\
-0.1	0.732675979183785\\
0.4	0.670117742590337\\
0.9	0.56695902605053\\
1.4	0.47270771712826\\
1.9	0.375910870219813\\
2.4	0.280649061902719\\
2.9	0.195207721970267\\
3.4	0.132738486967967\\
3.9	0.0801662759588319\\
4.4	0.0428482539739383\\
4.9	0.0210659499796657\\
5.4	0.00906149104046216\\
5.9	0.00338965527678557\\
6.4	0.00109098569520088\\
6.9	0.000287423037149039\\
7.4	6.39248807986355e-05\\
};
\addlegendentry{\footnotesize  MAP}

\addplot [color=red, line width=1.0pt, mark=x, mark options={solid, red}]
  table[row sep=crcr]{%
-1.6	0.905859750240154\\
-1.1	0.872021783526208\\
-0.6	0.824052409920449\\
-0.1	0.77228058285327\\
0.4	0.712797619047619\\
0.9	0.64537845705968\\
1.4	0.568679145794885\\
1.9	0.486591490571401\\
2.4	0.403133229792694\\
2.9	0.322345890410959\\
3.4	0.246139783893187\\
3.9	0.1786234641731\\
4.4	0.122814140707035\\
4.9	0.0783585261761855\\
5.4	0.0471924246476292\\
5.9	0.0257635454954121\\
6.4	0.012923726317085\\
6.9	0.00601045117082863\\
7.4	0.00248699475751676\\
7.9	0.000916309343319586\\
8.4	0.000288616217737294\\
8.9	8.10823593335189e-05\\
};
\addlegendentry{\footnotesize  SISO lifting}

\addplot [color=mycolor2, dashed, line width=1.0pt, mark=o, mark options={solid, mycolor2}]
  table[row sep=crcr]{%
-1.6	0.906796116504854\\
-1.1	0.86740692357936\\
-0.6	0.84366576819407\\
-0.1	0.803513071895425\\
0.4	0.796942375539004\\
0.9	0.738431151241535\\
1.4	0.673684210526316\\
1.9	0.620706575073602\\
2.4	0.549354131123568\\
2.9	0.479675546133284\\
3.4	0.408563535911602\\
3.9	0.336682907023955\\
4.4	0.274311198457992\\
4.9	0.208764246517518\\
5.4	0.153407134962767\\
5.9	0.104856279429334\\
6.4	0.0720049032538365\\
6.9	0.0450393091771719\\
7.4	0.0255550252534524\\
7.9	0.0136810934888591\\
8.4	0.00680433361787487\\
8.9	0.00291502998934674\\
9.4	0.00111309187259071\\
9.9	0.000372306628566283\\
10.4	0.000105975500507695\\
};
\addlegendentry{\footnotesize  Chase}

\addplot [color=mycolor3, dashed, line width=1.0pt, mark=asterisk, mark options={solid, mycolor3}]
  table[row sep=crcr]{%
-1.6	0.950381679389313\\
-1.1	0.938837920489297\\
-0.6	0.945111492281304\\
-0.1	0.918253079507279\\
0.4	0.913502109704641\\
0.9	0.892003297609233\\
1.4	0.877961234745154\\
1.9	0.833918715504265\\
2.4	0.788805970149254\\
2.9	0.738565782044043\\
3.4	0.686514886164623\\
3.9	0.633086079006383\\
4.4	0.563233208714709\\
4.9	0.501458312380569\\
5.4	0.410776523072633\\
5.9	0.333266693322671\\
6.4	0.264102196067174\\
6.9	0.192033463256964\\
7.4	0.136117982573621\\
7.9	0.0880606382465118\\
8.4	0.0533622681923229\\
8.9	0.0297261444856669\\
9.4	0.0149996421768438\\
9.9	0.00692233269577146\\
10.4	0.00276020914818501\\
10.9	0.000957779294136504\\
11.4	0.000286004499407452\\
11.9	7.15816252144156e-05\\
};
\addlegendentry{\footnotesize  Classical lifting}

\end{axis}
\end{tikzpicture}%
    \caption{FER  comparison of the Kerdock $\mathcal{K}[32,6]$ code decoded using MAP, SISO lifting, Chase and classical lifting decoders.}
    \label{fig:ferk32}
\end{figure}

SER performance of the $\mathcal{P}[128,120]$ code using the MAP decoder, the sub-optimal SISO decoder, the $\mathbb{Z}_4$ Chase decoder and the classical lifting decoder is shown in Fig. \ref{fig:serp128}. As expected, the MAP decoder has the best performance, while the sub-optimal SISO decoder is better then the classical lifting and the Chase decoder. The MAP decoder has a coding gain of about $1dB$ and $2dB$ compared to the Chase and the classical lifting decoders, respectively. The sub-optimal SISO decoder has a coding gain of about $0.5dB$ and $1.5dB$ compared to the Chase and the classical lifting decoders, respectively. Different decoding algorithms are also compared in terms of FER and results for the case of the $\mathcal{K}[32,6]$ code are shown in Fig. \ref{fig:ferk32}. The MAP decoder is optimal, so it has the best performance. We again notice that the sub-optimal SISO decoder is better than the classical lifting decoder and the Chase decoder. In the case of FER, the coding gain is more significant, and the MAP decoder achieves a coding gain of about $3dB$ compared to the Chase decoder.

\section{Conclusion}
\label{sc_fin}
This paper introduces two novel low complexity MAP decoding algorithms for decoding Kerdock and Preparata codes. The complexity of these decoding algorithms is $\mathcal{O}(N^2\log_2 N)$. A sub-optimal bit-wise soft-decision decoding algorithm, based on the decoder lifting technique, with complexity $\mathcal{O}(N\log_2 N)$ is also introduced. Compared to existing decoders, the novel decoders developed in this paper outperform existing ones in terms of error rate. Future work will deal with improving the error-correcting performance of the sub-optimal decoder, as well as developing new SISO decoders for other quaternary codes.


%

\appendices


\section*{Acknowledgment}

The authors are very grateful to the reviewers and the Editor, Prof. Li Chen, for their valuable comments that improved the quality and presentation of this paper.

\ifCLASSOPTIONcaptionsoff
  \newpage
\fi

\bibliographystyle{IEEEtran}
{\footnotesize
\bibliography{IEEEabrv,ref}}


\begin{IEEEbiography}[{\includegraphics[width=1in,height=1.25in,clip,keepaspectratio]{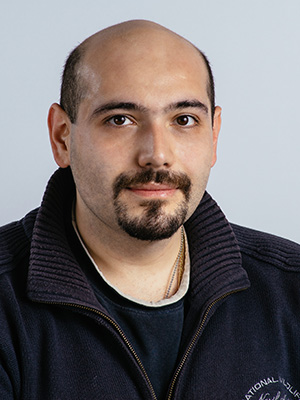}}]{Aleksandar Minja}
was born in Zrenjanin, Serbia, on June 19th 1987. He received the B.Sc (summa cum laude) and the M.Sc (summa cum laude) degrees in Electrical Engineering and Computer Science - Telecommunications Engineering and Signal Processing from the University of Novi Sad, Faculty of Engineering where he also finished his PhD and is currently working as an Assistant Professor. He is interested in channel coding and short packet communication.
\end{IEEEbiography}

\begin{IEEEbiography}[{\includegraphics[width=1in,height=1.25in,clip,keepaspectratio]{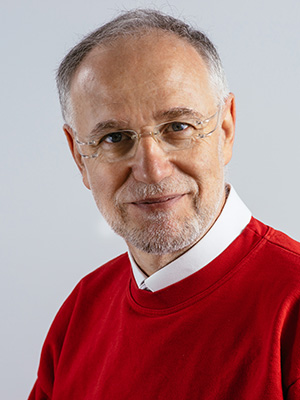}}]{Vojin Šenk}
is Professor of Information Theory and Encoding Techniques, as well as High-tech Entrepreneurship at the University of Novi Sad, Serbia. Born in 1958, he received his B.Sc degree in Electrical Engineering from the University of Novi Sad in 1981, M.Sc and PhD in Electrical Engineering from the University of Belgrade in 1989 and 1992, respectively. He is interested in the efficient decoding of channel codes and their implementation. 
\end{IEEEbiography}

\end{document}